\documentclass[12pt]{article}
\usepackage{amssymb}
\usepackage[english,dutch]{babel}
\parskip=\medskipamount

\begin{document}

 \raggedbottom
\addtolength{\baselineskip}{0.1\baselineskip}

\vspace*{1cm}  \noindent
\selectlanguage{english}
{\Large{\bf
Towards a Neo-Copenhagen Interpretation\\ of Quantum Mechanics}}

\vspace{1cm}

\noindent {\bf Willem M. de Muynck}\\
Theoretical Physics, Eindhoven University of Technology,\\ POB 513, 5600 MB Eindhoven,
The Netherlands \\
E-mail: W.M.d.Muynck@tue.nl\\

\noindent {\large \bf Abstract}

\noindent The Copenhagen interpretation is critically considered.
A number of ambiguities, inconsistencies and confusions are
discussed. It is argued that it is possible to purge the
interpretation so as to obtain a consistent and reasonable way to
interpret the mathematical formalism of quantum mechanics, which
is in agreement with the way this theory is dealt with in
experimental practice. In particular, the essential role
attributed by the Copenhagen interpretation to measurement is
acknowledged. For this reason it is proposed to refer to it as a
neo-Copenhagen interpretation.

\section{Introduction}\label{sec1}
Does quantum mechanics describe `just the phenomena', or does it
describe `reality behind the phenomena'? Fifty or sixty years ago,
under the influence of logical positivism/empiricism the majority
of physicists would presumably have given a positive answer to the
first question. Even today the influence of empiricism on the
interpretation of the mathematical formalism of quantum mechanics
is felt when using the term `observable' to refer to a physical
quantity represented by a Hermitian operator. And even though
neither Bohr nor Heisenberg can be reckoned a logical positivist,
their Copenhagen interpretation has been couched in the empiricist
language of `phenomena'. In particular, the emphasis placed on the
essential role played by the measurement arrangement has fostered
the idea that quantum mechanics is just dealing with phenomena to
be observed in the course of a measurement.

But, in the meantime there has been a remarkable change of
attitude. Following Einstein's criticism of the Copenhagen
interpretation, to the effect that ``Physics is an attempt
conceptually to grasp reality as it is thought independent of its
being observed'' (\cite{Einstein82a}, p.~81), there has been a
tendency to try to interpret quantum mechanics as a description of
an {\em objective} microscopic reality, `objective' to be
understood in the sense of `being independent of the observer
including his measuring instruments'. For instance, Bell
\cite{Bell90}: ``However the idea that quantum mechanics, our most
fundamental physical theory, is exclusively even about the results
of experiments would remain disappointing,'' and: ``To restrict
quantum mechanics to be exclusively about piddling laboratory
operations is to betray the great enterprise.'' Indeed, in quantum
mechanics textbooks of the last few decades hardly any reference
is being made at all to the measurement arrangement. Quantum
mechanics is presented as describing microscopic reality as it is
supposed to be independently of any measurement. Interpretations
like the many-worlds interpretation (Everett \cite{Ever73}) have
been especially devised so as to expel measurement from the
formulation of the theory. In the following I will refer to such
interpretations as `{\em objectivistic-realist} interpretations of
(the mathematical formalism of) quantum mechanics' (cf.
section~\ref{sec3}). It is the purpose of the present paper to
question the appropriateness of such objectivistic-realist
interpretations, and to develop an interpretation in which the
Copenhagen concern about the relation established by measurement
between the microscopic object and the macroscopic world, is duly
taken into account.

Unfortunately, the Copenhagen interpretation is far from
unambiguous. Thus, Bohr's view with respect to the state vector is
an {\em instrumentalist} one, in which the state vector is
considered as `just an instrument for calculating measurement
results', thus opposing the realist view of an electron as a wave
packet flying around in space. By contrast, even though Bohr
usually remained at a purely conceptual level, and refrained from
making ontological assertions, with respect to quantum mechanical
observables (``physical quantities'') his interpretation appears
to be a {\em realist} one (Folse (\cite{Folse85}), be it of a {\em
contextualistic} blend in which a physical quantity is
well-defined {\em only} within the context of the measurement
arrangement serving to measure that quantity. In this respect it
should be also noted that both Bohr and Heisenberg virtually
equate a `measurement result' (i.e. a value of a quantum
mechanical observable) with a property of the microscopic object,
possessed by the object {\em during} (Bohr) or {\em after}
(Heisenberg) the measurement. By von Neumann's projection
postulate this {\em contextualistic-realist} interpretation of
observables is extended to quantum mechanical states.

In the following sections I will first give a critical discussion
of a number of features of the Copenhagen interpretation. It will
be seen that Feyerabend's \cite{Feyerabend68} judgment that the
`Copenhagen point of view' is ``not a single idea but a mixed bag
of interesting conjectures, dogmatic declarations, and
philosophical absurdities'' is completely justified. Moreover, a
number of inconsistencies will be found. Nevertheless, the
tendency to dismiss as irrelevant the Copenhagen concern with
measurement is certainly unwarranted, and -in disagreement with
the often-heard assertion that the best explanation of the quantum
phenomena is that reality is objectively described by quantum
mechanics- not even plausible. On the contrary, the fact that all
knowledge we have about the quantum world is obtained by means of
measurement endorses Bohr's intuition that our view of the world,
laid down in our theories, must be colored by the interaction
between object and measuring instrument. Therefore it is necessary
to take this element of the Copenhagen interpretation very
seriously. It is important to note already here, however, that
this does not imply that it is possible to maintain a
contextualistic-realist interpretation in the Copenhagen sense
referred to above. Instead, an {\em empiricist} interpretation is
proposed, in which a quantum mechanical observable refers to the
measuring instrument rather than to the microscopic object (cf.
section~\ref{sec3}).

In order to avoid misunderstandings it should be stressed here
that in the following a `measuring instrument' is always taken to
be a material object the microscopic object is allowed to
physically interact with. The human observer and his consciousness
can be left out of consideration. They are assumed not to have any
physical influence on the measuring process after the measurement
arrangement has been set up. Actually, in a quantum measurement
the relation of a human observer to a measuring instrument is not
different from that in classical physics: his looking at the
(macroscopic) pointer of a quantum mechanical measuring instrument
will influence the measurement result (the pointer position) just
as little as it does in a classical measurement. Nowadays, quantum
measurements are often completely automated. The role of the human
observer may even be restricted to looking at the graphs produced
by his printer. Therefore in the following all allusion to `mind',
`consciousness', `free will', and `psychophysical parallelism'
will be ignored as being irrelevant to the subject. This implies
that certain issues, like von Neumann's `infinite regress',
`Wigner's friend', the `many-minds interpretation', and
`subjectivity' will not be discussed.

\section{Completeness of quantum mechanics}\label{sec2.1}
\subsection{Two notions of completeness}\label{sec2.1.1}
In the discussion between Einstein and Bohr on `completeness of
quantum mechanics' it is necessary to distinguish two different
notions of `completeness', to be referred to as `completeness in a
wider sense' and `completeness in a restricted sense',
respectively. The first notion is related to the impossibility of
subquantum theories, and can be formulated as follows:

\noindent {\em No subquantum theory can yield a more detailed
description of physical reality than is provided by quantum
mechanics} (completeness in a wider sense).

A reason to believe in `completeness in a wider sense' might be a
positivist fear of the metaphysical, rejecting hidden variables
because of their unobservable, and hence metaphysical, character.
As observed by Feyerabend \cite{Feyerabend68}, Bohr accepted von
Neumann's ``proof'' of the impossibility of hidden variables
completing the quantum formalism (\cite{vN32}, section~IV.2),
currently known to lack cogency (Bell \cite{Bell66}). Yet, this
was not the issue for Einstein when challenging Bohr's
`completeness' idea. The issue was rather different, viz.

\noindent {\em The quantum mechanical description of a particle
cannot be completed by determining precise values of position $\bf
r$ and momentum $\bf p$, because the disturbing influence of the
measuring instrument does not allow such a determination}
(completeness in a restricted sense).

The reason why Bohr thought quantum mechanics to be a complete
theory was not any fear of the metaphysical. Instead, it is the
existence of the `quantum of action', represented by Planck's
constant $h$, which is responsible for the impossibility of a
simultaneous precise determination of position and momentum of a
particle. For Bohr the concomitant impossibility of having a
vanishing interaction between microscopic object and measuring
instrument implies the impossibility of a sharp distinction
between these objects, which therefore constitute an indivisible
whole, manifesting itself as a `quantum phenomenon' (this is known
as Bohr's {\em quantum postulate}, e.g.~\cite{Bohr49}). Due to
this circumstance position and momentum are each defined with a
certain latitude, which latitudes satisfy the Heisenberg
uncertainty relation. This issue is characteristic of quantum
mechanics itself, and is quite independent of the question of
hidden variables (which transcends the domain of quantum
mechanics). It is `completeness in a restricted sense' that is the
issue in the discussion between Bohr and Einstein on the so-called
`thought experiments', culminating in the Einstein-Podolsky-Rosen
(EPR) proposal \cite{EPR}, which should be seen as an ultimate
attempt from Einstein's part to prevent an appeal by Bohr to the
measurement interaction in countering his objections.

Viewed in this way, the discussion between Einstein and Bohr on
the completeness of quantum mechanics is not at all about whether
quantum mechanics is a `theory of everything', not to be surpassed
by any other more detailed theory. It is rather about the
interpretation of quantum mechanics, viz. whether a {\em realist}
interpretation of quantum mechanical observables can be either
{\em objectivistic} (Einstein) or must be {\em contextualistic}
(Bohr). It was Einstein's conviction that a sound physical theory
must yield a description of an {\em objective} reality, and not
merely of a reality that is in interaction with an observer, or
even with a measuring instrument. The moon is there when nobody
looks. The properties of matter, like electric conductivity,
radioactivity, etc., do not seem to be dependent on our
observations; we should try to devise theories describing these
properties as being independent of any observation. This idea is
at the basis of the notion of `element of physical reality',
introduced in the EPR paper \cite{EPR}, corresponding to a
physical quantity the value of which can be predicted with
certainty {\em without in any way disturbing the system} (cf.
section~\ref{sec2.4}). By Bohr \cite{Bohr35} the unambiguity of
this notion was challenged precisely because of its {\em
non}-contextuality, ignoring that the (whole) experimental arrangement
must be taken into account when discussing quantum phenomena.

It is important to stress here that by EPR an element of physical
reality was presented as a {\em quantum mechanical measurement
result}, which, for this reason, could be equated by Bohr to a
`quantum phenomenon'. Indeed, the discussion took place completely
within the domain of quantum mechanics. No hidden variables,
hence, no `completeness in a wider sense' were involved. As far as
hidden variables might be thought to be involved, the question
just is whether the quantum mechanical observables themselves can
play the roles of hidden variables, in the sense that to a
particle a well-defined value of an observable can be attributed,
possessed prior to, and independently of measurement. However,
this is a matter of the interpretation of quantum mechanics
(either objectivistic-realist or contextualistic-realist) rather
than a fundamental change to a more encompassing theory. It should
be borne in mind that this focussing of the attention on {\em
quantum mechanical} notions has severely restricted the scope of
physical reasoning. Thus, in this way the possibility is not taken
into account that Bohr might be right in attributing a contextual
meaning to quantum mechanics, but that Einstein's `element of
physical reality' might be given an unambiguous meaning when
viewed as a {\em non}-quantum mechanical concept. By disregarding
the latter possibility it became possible that Bohr's victory with
respect to the issue of `completeness in a restricted sense' was
generally misinterpreted as a victory with respect to
`completeness in a wider sense', denouncing Einstein's
introduction of `elements of physical reality' as a metaphysical
endeavor.

The idea that quantum mechanics might be `complete in a wider
sense' is a very unusual one, not applicable to any of the
physical theories developed in the past. Therefore it is not clear
at all why we might want to apply it to quantum mechanics.
Admittedly, this theory has a very large domain of application,
and it is not clear under which physical circumstances its
boundaries might come into sight. But this was true as well for
the theory of classical mechanics, including Maxwell's field
theory, which encompasses virtually all of physics known $100$
years ago (compare Kelvin \cite{Kel}). Nevertheless, $20^{th}$
century physics has been dominated by quantum mechanics and
relativity theory, and the classical theory has been recognized to
be valid only if masses are sufficiently large and velocities are
small compared to the velocity of light. Quantum mechanics does
not seem to be so well understood that something analogous could
not be expected (for instance, for measurement processes
monitoring very short times, cf. section~\ref{sec4}).

The idea that quantum mechanics is `complete in a wider sense' can
hardly be attributed to either Einstein, Bohr or Heisenberg, who
were open-minded to the possibility that quantum mechanics may
have to be superseded by still more advanced theories. If the idea
of `completeness in a wider sense' can be attributed to the
Copenhagen interpretation at all, it is because this
interpretation is not a consistent theory, but a set of ideas
stemming from different sources. One such source is the empiricist
philosophy of science which was dominating the first half of the
$20^{th}$ century. In a certain fundamentalist form (often
referred to as anti-realism) the empiricist philosophy not only
advocated to be very cautious about theoretical concepts not based
on observation, but it even declared objects corresponding to such
concepts to be non-existent (for instance, atoms, the world
aether). If taken seriously, application of this advice to atoms
would presumably even have been an impediment to their
experimental discovery. Although it may not be advisable to assume
the physical existence of every theoretical notion we may be able
to think of, the opposite attitude does seem to be equally
unproductive. As a means of promoting science, `fear of the
metaphysical' has its boundaries. Sometimes a leap of imagination
may be advantageous.

Up to now our experience has been that behind the phenomena
described by a certain theory there is a wealth of new physics to
be described by more penetrating theories. Thus, the behavior of a
billiard ball as observed by a billiard player is adequately
described by the classical theory of rigid bodies. However, we
need solid state physics to take into account its atomic
constitution. By analogy, it would be rather frivolous to assume
that quantum mechanics is the ``theory of everything'', never to
be superseded by more encompassing (subquantum) theories. Even
though at this moment we do not have any experimental indication
with respect to the boundaries of the domain of application of
quantum mechanics (at least in its generalized form, cf.
section~\ref{sec2.8}), the idea of `completeness in a wider sense'
would be inappropriate from a methodological point of view. As a
matter of fact, it is by now well known that the standard
formalism of quantum mechanics, as presented in quantum mechanics
textbooks, is not able to describe all measurements possible
within the quantum mechanical domain (e.g. de Muynck
\cite{dM2002}). For this reason, at least textbook quantum
mechanics cannot be `complete in a wider sense'. It will not be
demanded in developing a neo-Copenhagen interpretation of quantum
mechanics.

\subsection{Objectivity and contextuality}\label{sec2.1.2}
There is yet another lesson to be learned from the billiard ball
analogy. It is widely assumed that, in contrast to quantum
mechanics, classical mechanics yields an {\em objective}
description of reality, not needing any reference to observation.
From the example we see that this is not generally true. Rigidity
is not an {\em objective} property of a billiard ball. Due to its
atomic constitution, a billiard ball behaves as a rigid body only
under certain circumstances. If hit hard enough it may start to
vibrate, and may even split. Hence, the theory of rigid bodies is
applicable only within a certain domain of experimentation, and,
therefore, has only a {\em contextual} meaning, quite analogous to
Bohr's view on the contextual meaning of quantum mechanical
observables. As far as Einstein's idea of the desirability of an
{\em objective} description might have its origin in a classical
paradigm, this seems to be unwarranted. Moreover, it is evident
that, analogous to Bohr's contention with respect to quantum
mechanical observables, for a rigid body description of a billiard
ball the context is determined by the {\em whole experimental
arrangement}. This implies that `rigidity of a billiard ball' is a
{\em contextual} property even if no observation is made at all.
Einstein's requirement of objectivity of the quantum mechanical
description might be justified if quantum mechanics were `complete
in a wider sense'. This seems to be too strong a requirement,
though.

However, Einstein's requirement of `objectivity' is not related to
the issue of `completeness in a wider sense'. Indeed, his quarrel
with Bohr was about `completeness in a restricted sense'. It was
unacceptable to Einstein that properties of a world, existing
independently of the observer, would depend on its being observed.
Indeed, it would be preposterous to assume that the rigidity of a
billiard ball is dependent on its being looked at, or that the
moon would not exist if it is not observed by any observer.
The billiard ball example can teach us how this conundrum can be
solved, both for classical and quantum mechanics. For this purpose
it is advantageous to assume `incompleteness in a wider sense' of
these theories.

It is important to take into account the physical reason why a
billiard ball is rigid, even if it consists of vibrating atoms.
This reason, of course, is that, due to the tight bindings between
the atoms within the ball, the vibrations are so small that they
are not observed at the macroscopic level of observation to which
the theory of rigid bodies is applicable. We should distinguish
`reality' from our `description of reality'. A billiard ball is
not ``really'' a rigid body; we can only describe it as one as
long as it behaves accordingly. Hence, as far as rigidity is a
property of the ball, it is a {\em contextual} property. Taking
into account atomic vibrations makes this clear. Moreover, it
demonstrates that for a description of these vibrations we need a
theory different from rigid body theory (for instance, an atomic
theory of the solid state). The contextual meaning of rigid body
theory is evident because the concept of `rigidity' loses its
meaning when the atomic vibrations are no longer negligible.
Actually, we should distinguish two concepts of `rigidity', the
`rigidity' concept of rigid body theory being quite different from
that of an atomic theory of the solid state. An interesting
question is also whether, in contrast to the concept of
`rigidity' of rigid body theory, the atomic theory does yield an {\em
objective} description of atomic vibrations, or whether under
certain conditions these vibrations may lose their meaning too.
However, this subject will not be pursued here any further.

For quantum mechanics the situation may be analogous. {\em
Submicroscopic} (hidden variables) concepts of position and
momentum may be different from the corresponding quantum
mechanical concepts. Bohr may be right that, like rigidity,
quantum mechanical observables too have a contextual meaning only.
If so, a subquantum theory will be necessary to yield the
objective description Einstein aspired to. Possibly, Einstein's
requirement that quantum mechanics itself yield such a description
is unnecessarily asking too much of this latter theory. We have
some indications that this, indeed, may be the case. Thus,
although at the time of the Bohr-Einstein discussion it was an
open question whether values of quantum mechanical observables can
be attributed to the microscopic object as objective properties
(this possibility was denied by the Copenhagen interpretation on
dubious grounds, cf. section~\ref{sec2.1.3}), by the
Kochen-Specker theorem and its generalizations
\cite{KS,Mermin90,Mermin90b,Peres91,Mermin93}, as well as by
certain derivations of the Bell inequalities (e.g. \cite{dM2002},
section~9.4.1), we are convinced now that such an attribution is
impossible. Hence, in general it is impossible to assume that a
free particle had a well-defined value of quantum mechanical
momentum prior to measurement. This is one of the basic tenets of
the Copenhagen interpretation, advocated in particular by Jordan
\cite{Jord34}, but generally felt to characterize this
interpretation's view on the issue of {\em (in)determinism}, which
(indeterministic) view -as is well known- was rejected by
Einstein. His assertion that God does not play dice may be
interpreted as expressing a conviction that in an ideal
measurement an observation of a quantum mechanical measurement
result can be explained because the observable had its value prior
to the measurement. This is sometimes referred to as a principle
of `faithful measurement' (e.g. Redhead \cite{Redhead}). It seems
that at least on the issue of the objectivity of quantum
mechanical measurement results Bohr can be granted a victory over
Einstein, even though it is questionable whether this extends to
the issue of (in)determinism.

\subsection{To explain or not to explain}\label{sec2.1.3}
The distinction between `quantum mechanical measurement results'
and `subquantum elements of physical reality' will play an
important role in my attempt to free the Copenhagen interpretation
from a number of features reducing its trustworthiness. Jordan's
`creation-out-of-the-blue' philosophy is one of these. The
Copenhagen abandonment of explanation of a measurement result by
referring to a property of the microscopic object, possessed prior
to measurement (for instance, according to Jordan \cite{Jord34}
before a position measurement a particle is ``neither here,
neither there'') has been a bone of contention to those believing,
with Einstein, that a decent physical theory must tell us
something about objective reality. Indeed, the Copenhagen idea of
`indeterminism' is based on a conviction that quantum mechanics
cannot be completed by means of so-called hidden variables. This
conviction is based on von Neumann's ``proof'' of the
impossibility of the existence of hidden variables (\cite{vN32},
section~IV.2), which, however, was demonstrated by Bell
\cite{Bell66} not to be cogent (see also Feyerabend
\cite{Feyer56}). Hence, it is possible that here the Copenhagen
interpretation underestimates the capacity of quantum mechanics to account for certain features of reality.
Perhaps Heisenberg does
not even follow Jordan all the way when he interprets his
uncertainty relation as referring to mutual disturbance of
position and momentum in a simultaneous measurement of these
observables. Indeed, this seems to imply that an
observable is not disturbed by an ideal measurement of that very observable, suggesting that, in agreement with the principle of `faithful measurement', prior to measurement the observable must have had the same value it has after the measurement. Even though,
due to the Kochen-Specker and Bell theorems, this cannot be
literally true, Heisenberg's intuition might be correct in the
sense that, contrary to Jordan's contention, a quantum mechanical
measurement may yield evidence on a previously existing
`element of physical reality', which cannot be described by
quantum mechanics, however, but which has to be described by a
subquantum theory. Note, too, that Bohr has repeatedly warned
against the idea that during measurement a quantum mechanical
measurement result is created (as a property of the microscopic
object).

Measurements of correlations between two observables may provide
even more pressing arguments for the existence of `subquantum
elements of physical reality'. Consider, for instance, a free
particle, for which the momentum observables $P(t)$ and $P(t')$
commute for all values of the times $t$ and $t'$. It, therefore,
follows from the mathematical formalism of quantum mechanics that
two consecutive measurements of momentum of a free particle will
yield the same measurement result. The perfect correlation between
the results of two consecutive momentum measurements could be
explained in a natural way by conservation of momentum, combined
with the fact that, due to the commutativity of the momentum
operators $P(t)$ and $P(t')$, the measurements of these
observables do not mutually disturb. The perfect correlations of
the measurement results would be completely unexplained if they
were not reducible to some feature of the microscopic reality
behind the measurement phenomena. A desire to explain a similar
perfect correlation of quantum mechanical observables of two
different particles in the EPR experiment may be responsible for
the idea that, if the measured quantities cannot be objective
properties of the particles, possessed prior to measurement,
nonlocal influences between distant measurements must be involved.

Even though values of Hermitian operators (analogous to the
rigidity concept of rigid body theory) cannot play the roles of
`elements of physical reality', this does not imply that they
would not refer to some aspect of reality that is actually probed
by the measurement of the corresponding observable. Thus, quantum
mechanical momentum of a particle may refer to some subquantum
notion of `momentum', related to the quantum mechanical notion in
a way analogous to the relation between the notion of `rigidity'
of rigid body theory and its representation in atomic solid state
theory: conservation of momentum may be compared to conservation
of the spherical shape of the surface of a billiard ball caused by
the (approximate) preservation of the relative positions of the
atoms; whereas under standard conditions these features strictly
correspond to our observations, a description by the more detailed
theory may account for deviations which become important under
non-standard conditions.

Einstein's reluctance to grant completeness to quantum mechanics
was not induced by any preference for the determinism of classical
mechanics, but rather by the idea that our physical theories must
tell us something about the world as it exists objectively and
independently of measurement. Although this ideal may not be
applicable to all physical theories (e.g. rigid body theory), and,
in particular, not to quantum mechanics, it is not too far-fetched
to assume with Bohr that quantum mechanics at least may tell
something about a {\em contextual} reality that is in interaction
with a measuring instrument.
Nor does it mean that no other explanations can be attempted. As a
matter of fact, within the Copenhagen interpretation such an
attempt is made by invoking von Neumann's projection postulate,
trying to explain the strict correlation between consecutively
measured values of momentum by assuming that a momentum
measurement projects the quantum state into an eigenstate of the
measured observable. However, von Neumann's projection postulate
is inapplicable to most measurement procedures (e.g. de Muynck
\cite{dM2002}, chapter~3). Moreover, as an explanation it does not
seem to be very plausible if compared with the possibility that
`conservation of momentum' tells us something about a property of
the object that is conserved while it is {\em not} interacting
with a measuring instrument (an `element of physical reality'). It
seems evident that, if quantum mechanical observables have a
contextual meaning only, such a property, if existing, must have a
{\em non}-quantum mechanical nature.

Of course, it is possible to refrain from any explanation of
correlations of consecutive quantum mechanical measurement results
like the one discussed above, or of the strict (EPR) correlations
obtained in a simultaneous measurement of the $z$-components of
the spins of a spin-$1/2$ particle pair prepared in the singlet
state. This would suit a strictly empiricist view of quantum
mechanics (e.g. van Fraassen \cite{vF91}), in which it is deemed
sufficient that a physical theory just describes the phenomena,
without any necessity to explain them. As is evident from its
adoption of von Neumann's projection postulate, the Copenhagen
interpretation does not hold to this strictly empiricist view. It
does not refrain from explanation. On the contrary, it seems to be
so inclined toward explanation that it is ready to accept a
suspect procedure like von Neumann's projection postulate as a
means of explaining correlations in consecutive or in joint
measurements. It seems to me that, in devising a neo-Copenhagen
interpretation of quantum mechanics, we should maintain the demand
that physical theories provide explanations, but we must accept
that, like rigid body theory, also quantum mechanics may not
explain everything. In particular, it does not explain why a
certain measurement result is obtained when a measurement of a
quantum mechanical observable is carried out. For such an
explanation, if it exists, we will have to resort to more detailed
(subquantum) theories, much in the same way rigidity of a billiard
ball is explained by the tight bindings between the atoms within
the ball, to be described by a sub-rigid body theory. If quantum
mechanics is not `complete in a wider sense', we may expect that
subquantum theories will once be found. Far from being
methodologically objectionable, subquantum (hidden variables)
theories may be illuminating by allowing different levels of
discourse.

\section{Individual-object (individual-particle) versus ensemble interpretation of
the wave function} \label{sec2.2}

The question of whether quantum mechanics is `complete' is often
formulated in terms of the wave function or state vector. Einstein
was very clear at that. According to him ``the $\psi$-function is
to be understood as the description not of a single system but of
an ensemble of systems'' (Einstein \cite{Einstein82}, p.~671). He
thought that our inability to yield a more precise description
than a statistical one is a consequence of our ignorance about the
precise values of position and momentum of a particle. According
to the Copenhagen interpretation such an ensemble interpretation
is not suitable for quantum mechanics, because it is in
disagreement with the idea of `completeness in a restricted sense'
(cf. section~\ref{sec2.1.1}), disallowing an object to
simultaneously have well-defined values of position and
momentum\footnote{Here it is temporarily ignored that in the
Copenhagen interpretation a quantum mechanical measurement result
is generally interpreted as a {\em post-}measurement property of
the object. The Copenhagen lack of distinction between pre- and
post-measurement properties will be discussed in
section~\ref{sec2.7}.}. According to the Copenhagen interpretation
the wave function must be seen as a fundamentally probabilistic
(as opposed to statistical) description of an {\em individual}
object, allowing for the essential {\em in}determinism which is a
consequence of `completeness in a restricted sense'. The
difference between the Copenhagen individual-particle
interpretation and Einstein's ensemble one marks their
fundamentally different attitudes with respect to explanation of
quantum mechanical measurement results.

Sometimes the distinction between an ensemble interpretation and
an individual-object interpretation is characterized in terms of
the distinction between an {\em epistemic} and an {\em ontic}
interpretation, respectively (e.g. Primas \cite{Primas83}),
juxtaposing `description of our knowledge' to `description of
reality'. It seems to me, however, that this is a potentially
misleading distinction, introduced in the first place to discredit
an epistemic interpretation as being part of psychology, not
physics. It should be noted, however, that, independently of its
interpretation, a physical theory is a representation of our
knowledge (about some part of reality), and, therefore, is always
epistemic. This holds true for the wave function as well,
independently of whether it is considered as a description of an
individual particle or of an ensemble. Contrary to what is
intimated by an ontic interpretation, an electron is not a wave
function flying around in space. Electrons belong to physical
reality, wave functions can be found in quantum mechanics
textbooks.

I will not discuss here any further the question of
whether the Copenhagen interpretation is ontic in the sense given
above, because it seems to me that different adherents to this
interpretation may give different answers (often couched in a
terminology referring to the objectivism/subjectivism dichotomy).
For instance, it seems evident that Bohr's (instrumentalist)
interpretation of the wave function is not ontic. However, as
already remarked in section~\ref{sec1}, there has been a strong
tendency towards an ontic interpretation, considering a wave
function to describe the reality of an individual microscopic
object, much in the same way as the theory of rigid bodies seems
to describe the reality of a billiard ball (see also
section~\ref{sec3}). Nowadays the `epistemic' terminology has
become acceptable again by stressing that the wave function must
be seen as representing {\em information} which is available about
the object. This would be a step forward if the experimental means,
necessary to obtain this information, are not left out of
consideration (which is seldom the case, however).

Unfortunately, reliance on the distinction `epistemic versus
ontic' has had a considerable impact. Thus, in the distinction
between `statistical' and `probabilistic' interpretations of the
quantum mechanical formalism it has been assumed that the latter
interpretation implies {\em in}determinism in an {\em ontic} sense
rather than in the epistemic sense of lack of knowledge. It must
be stressed, however, that we do not have any hard experimental
evidence for such a contention. The arguments advanced by the
Copenhagen and other interpretations in favor of an
individual-particle interpretation of the wave function are based
on an unwarranted lack of distinction between microscopic reality
itself and its quantum mechanical description. For this reason I
will not consider the epistemic/ontic dichotomy any further, but I
will focus here on the individual-particle/ensemble dichotomy in
which a quantum mechanical wave function may refer to either one
of two different objects, both ``ontically'' existing in reality,
viz. an {\em individual} object or an {\em ensemble} of such
objects, realizing that in both cases the quantum mechanical
description is an approximate one.

One prominent argument in favor of an ensemble interpretation of
the wave function or state vector is the fact that
quantum mechanical probabilities can be experimentally approached
only by repeating an experiment a large number of times, and by
determining relative frequencies of measurement outcomes. The
possibility of an individual-particle interpretation then hinges
on the question of whether all individual preparations in an
ensemble are `identical preparations', not allowing any
specification by which they could be distinguished. By Von Neumann
\cite{vN32} this was taken up as signifying {\em homogeneity} of
an ensemble described by a wave function (in contrast to the
apparent {\em in}homogeneity of a von Neumann ensemble described
by a non-idempotent density operator). I will now discuss two
reasons why I think that an individual-particle interpretation of
the wave function is undesirable, even though at this moment this is
not strictly falsifiable due to lack of any possibility to
distinguish, other than by the measurement results obtained in
quantum mechanical measurements, between individual realizations
of members of an ensemble described by a wave function.

As a first reason, let us consider a particle impinging on a
double-slit system. Nowadays such experiments are no longer
`thought experiments'. For instance, in neutron interferometry
such experiments are routinely performed (e.g.
\cite{WeKl,SuRaTu,Zeil86}). In an individual-particle
interpretation such experiments immediately entail a conceptual
problem. As a matter of fact, the wave function is split into two
parts, each corresponding to one of the slits. This is sometimes
interpreted as `the particle going through both slits at the same
time'. A more appropriate interpretation might be thought to be
the Copenhagen one, to the effect that within the experimental
arrangement the particle concept is not well-defined, and a `wave'
terminology should be used to interpret what is going on
(particle-wave duality). Unfortunately, the Copenhagen idea of
particle-wave duality, asserting that, depending on the
experimental arrangement, a quantum mechanical object ``is''
either a particle or a wave, must be considered obsolete by now.
In interference experiments {\em both} particle and wave aspects
can be observed within one and the same experimental arrangement.
For instance, in interference experiments with electrons and
neutrons it is possible to see a gradual development of the
interference pattern, built up by local (particle-like) impacts
exhibiting particle-like and wave-like behavior in one and the
same measurement arrangement (compare figure~4.4 of de Muynck
\cite{dM2002}). This experimentally demonstrates that in
interference experiments the wave aspect of the phenomenon cannot
be explained by considering an individual particle as a wave
(which, incidentally, would be analogous to explaining rigidity of
a billiard ball by means of a model of closely packed rigid
atoms). For Bohr this has been occasion to revise his views, to
the effect that electrons, protons and neutrons are considered by
him to be always particles, whereas light would always be a wave,
photons being artefacts of the quantum mechanical description (cf.
Murdoch \cite{Murd}).

Indeed, it is consistent with all experimental evidence to assume
that electrons and neutrons maintain a particle-like behavior
while passing through the interfero-\linebreak meter. If detectors
are placed within the interferometer, the particle is always found
in {\em one} of the possible paths, never in both paths at the
same time. Something similar is observed when a photon is allowed
to impinge on a semi-transparant mirror (hence, it seems that in
this respect Bohr's revised view is questionable as well). Whereas
the wave function is split into approximately equal parts
(transmitted and reflected parts), the photon travels one way or
the other, as can be observed by putting photon counters in each
of the outgoing directions. Evidently, if the wave function were
interpreted as a description of a single particle or photon, the
object (wave?) would be equally in both paths. This does not seem
to be in agreement with the results of optical experiments,
however.

In the Copenhagen probabilistic interpretation of the wave
function a particle ontology is employed, but it is thought to be
undefined whether the particle is in one path or the other; it
might even jump to and fro between the paths. Analogously to the
case of conservation of momentum, discussed above, we have no
reason to believe that in a position measurement von Neumann
projection is the reason that we find the object in one and the
same beam when consecutive position measurements are carried out.
For an explanation of this it is far more reasonable to appeal to
inertia as a general property of matter, valid in the microscopic
domain as well (although, possibly, not in the classical
mechanical sense). The reason of the assumption of indeterminism
by the Copenhagen interpretation might be also a consequence of an
ill-founded idea of `completeness of quantum mechanics in a wider
sense', precluding a distinction between quantum mechanical
measurement results and (non-quantum mechanical) properties of the
microscopic object. Einstein's statistical (ensemble)
interpretation of the wave function is no less in agreement with
experiment, and its determinism in the sense of `faithful
measurement' seems to be quite a bit closer to what should be
expected from a physical theory. For this reason it seems to be
more promising to try to remedy shortcomings of this `ensemble'
approach than to keep trying to circumvent the paradoxes going
with an individual-particle interpretation. One of these paradoxes
is discussed in the next section. It yields a second reason to
doubt the possibility of an individual-particle interpretation of
the wave function, if the particle is considered in von Neumann's
sense as a member of a homogeneous ensemble.

\section{(In)homogeneity of ensembles}\label{sec2.3}
When the idea of `homogeneity of an ensemble
described by a pure state' is applied to an entangled state
\begin{equation}
| \psi_{12} \rangle = \sum_{ij} c_{ij} | \alpha_{1i} \rangle |
\beta_{2j} \rangle,\;c_{ij}\neq c_{1i}c_{2j}, \label{1}
\end{equation}
$\{| \alpha_{1i} \rangle\} $ and $\{| \beta_{2j} \rangle\} $
orthonormal sets in the Hilbert spaces of particles $1$ and $2$,
respectively, we are confronted with a consistency problem. By
taking a partial trace of the density operator $\rho_{12} =|
\psi_{12} \rangle\langle  \psi_{12}|$, the state of particle $1$
is obtained as
\begin{equation}
\rho_1=Tr_2  \rho_{12} =  \sum_ j r_j |\phi_{1j} \rangle\langle
\phi_{1j}|, \;|\phi_{1j} \rangle = \sum_i
\frac{c_{ij}}{r_j^{1/2}}|\alpha_{1i}\rangle, \; r_j =\|\sum_i
c_{ij}|\alpha_{1i}\rangle\|^2.\label{2}
\end{equation}
The important point is that $\rho_1$ does not describe a pure
state but a von Neumann ensemble, which, allegedly, is
inhomogeneous. However, since particle $1$ is a subsystem of the
two-particle system described by the pure state (\ref{1}), we do
not have any means to distinguish between different members of the
(allegedly homogeneous) ensemble, {\em also} if only one of the
particles is considered. Hence, it would seem that the particle
$1$ ensemble must be homogeneous too.  This poses the question of
whether $\rho_1$, given by (\ref{2}), describes a homogeneous or
an inhomogeneous ensemble, the latter answer implying an
inconsistency if the pure state (\ref{1}) would correspond to a
homogeneous one.

The question cannot be answered straightforwardly in an
unambiguous way. As a matter of fact, for every density operator
of the type (\ref{2}) an entangled state of the form (\ref{1}) can
be constructed such that the density operator can be obtained by
taking a partial trace. If pure states are homogeneous, this would
seem to imply that ensembles of the type described by (\ref{2})
must be homogeneous too. This is one way to try to prevent the
above-mentioned inconsistency: consider all quantum mechanical
ensembles, either pure states or mixtures, as homogeneous. This is
actually done in the {\em minimal} interpretation advocated by
e.g. Park and Band \cite{Park73,BaPa76}. In this way a problem of
von Neumann's interpretation, caused by the non-uniqueness of the
decomposition of the ensemble described by density operator
(\ref{2}) into distinct subensembles,
might be solved. However, there is a different solution, viz., the
one advocated by Einstein, considering all ensembles, both pure
states as well as mixtures, as {\em in}homogeneous. Hence, on the
basis of the present considerations no definite answer can be
given.

In order to discuss the problem of (in)homogeneity more fully, I
will assume that in a homogeneous ensemble a quantum mechanical
measurement produces a {\em random} sequence of measurement
results which is homogeneous too. Here `homogeneity of a random
sequence' is taken in the sense of von Mises' theory of random
sequences \cite{Mises} according to the definition:

\noindent {\em A random sequence is homogeneous if every allowed
subsequence has the same relative frequency as the original
sequence} (homogeneity of a random sequence).

\noindent According to von Mises a subsequence is allowed if its
members are selected using an algorithm that does not depend on
the values of the selected members (for instance, in a random
sequence consisting of $0$'s and $1$'s, the criterion `select all
$0$'s' is not allowed). Now the question of whether an ensemble,
described by density operator $\rho= \sum_j r_j |\phi_j
\rangle\langle \phi_j|$ is inhomogeneous, hinges on the question
of whether there exists an allowed selection procedure to select
subensembles described by state vectors $|\phi_j \rangle$ (which
subensembles in general yield distinct measurement results if the
state vectors are different for different $j$). As argued above,
von Neumann's choice to answer this question affirmatively is less
justified than it appear to be at first sight.

By d'Espagnat (\cite{EspCFQM}, chapter~7.2) a distinction has been
drawn between {\em proper} and {\em improper} mixtures, described
by the same density operator. In the case of a proper mixture,
represented by density operator $\rho= \sum_j r_j |\phi_j \rangle
\langle \phi_j|$, an allowed selection procedure exists, viz.
selection on the basis of the parameter setting $j$ of the
preparation apparatus corresponding to the preparation of the
subensemble described by $|\phi_j\rangle$. This selection
procedure is allowed because the parameter setting $j$ determines
the state $|\phi_j \rangle$ rather than the other way around.
According to the von Mises criterion a proper mixture must
therefore be inhomogeneous.

When we apply the above reasoning to improper mixtures (obtained
from a pure entangled state by partial tracing) the outcome is
less unambiguous. The question is whether also in the case of an
improper mixture an allowed selection procedure can be found
yielding a subsequence with relative frequency differing from the
one obtained from $\rho$. Using (\ref{1}) and (\ref{2}) (taking
$\rho_1 =\rho$ and $\phi_{1j}=\phi_{j}$), a selection procedure
could be contemplated on the basis of the existence of a polar
decomposition (Schmidt \cite{Schm}), to the effect that any
two-particle state $| \psi_{12} \rangle$ can be written according
to
\begin{equation}
| \psi_{12} \rangle = \sum_i c_i | \alpha^{(s)}_{1i} \rangle |
\beta^{(s)}_{2i} \rangle, \label{3}
\end{equation}
where $\{|\alpha^{(s)}_{1i}\rangle\}$ and
$\{|\beta^{(s)}_{2i}\rangle\}$ are special orthonormal sets of
eigenvectors of observables of particles $1$ and $2$,
respectively, determined by the state vector $|\psi_{12}\rangle$
(these vectors turn out to be the eigenvectors of the reduced
density operators $\rho_1$ and $\rho_2$, obtained from
$|\psi_{12}\rangle\langle \psi_{12}|$ by partial tracing). Hence,
for any (improper) mixture there exist two observables,
$A^{(s)}_1$ and $B^{(s)}_2$ (having the vectors
$|\alpha^{(s)}_{1i}\rangle$ and $|\beta^{(s)}_{2i}\rangle$ as
eigenvectors, respectively) which are strictly correlated
according to (\ref{3}). Now a measurement of $B^{(s)}_2$ on
particle $2$ might seem to yield an allowed selection procedure
for selecting subensembles of particle $1$ described by the
vectors $|\alpha^{(s)}_{1i}\rangle$. Since the relative
frequencies of a measurement performed on particle $1$ will in
general be different for different values of $i$, the ensemble
corresponding to $\rho=\rho_1$ (in which the measurement results
for particle $2$ are ignored) would then be inhomogeneous.
However, as will be seen next, there is a catch to this argument.

Another way to look at this problem is by not only considering a
measurement of observable $B^{(s)}_2$ on particle $2$ but also a
simultaneously performed measurement of observable $A^{(s)}_1$ on
particle $1$, yielding two (correlated) sequences of measurement
results. Due to the correlation, subsequences of measurement
results of particle $1$ can be selected, conditional on certain
measurement results of particle $2$. Once again, the question is
whether this selection procedure is an {\em allowed} one. The
answer to this question hinges on the (in)dependence of the value
of observable $B^{(s)}_2$ (constituting the criterion of
selection) with respect to the value of observable $A^{(s)}_1$.

In general, measurements performed on two different particles can
be considered to be independent. This suggests that the selection
procedure is an allowed one. However, we do not have a general
situation here. With respect to the measured observables state
vector (\ref{3}) is a very special one, warranting a strict
correlation between the measurement results of the two particles.
The selection procedure makes an essential use of this strict
correlation, even to the effect that occasionally a measurement on
one particle is interpreted as a measurement of the correlated
observable of the other particle (compare the EPR experiment to be
discussed in section~\ref{sec2.4}). If this is taken into account,
the selection procedure does not seem to be allowed any more,
because in this experiment selection of a measurement result of
particle $2$ is equivalent to selection of the corresponding
measurement result of particle $1$, thus making the selection
procedure depend on the value of the selected measurement result.
On this basis homogeneity of improper mixtures could be thought to
be maintainable.

However, this conclusion is changed if we take into account the
possibility of measuring, in coincidence with $B^{(s)}_2$, an
observable $A'_1$ of particle $1$ differing from $A^{(s)}_1$. In
general, in the state (\ref{3}) observables $A'_1$ and $B^{(s)}_2$
are not strictly correlated, and their measurement results can be
considered as independently obtained. This implies that, according
to von Mises' definition, the ensemble represented by (\ref{2})
can be considered as {\em in}homogeneous. Homogeneity of mixtures
(either proper or improper) seems to be maintainable only if the
experimental possibility to split quantum mechanical ensembles
into distinct subensembles is ignored. We will see in the
following that this issue is closely related to the Copenhagen
negligence of the difference between the notions of `preparation'
and `measurement', causing a strong focussing on measurement of
the special observables $A^{(s)}_1$ and $B^{(s)}_2$ if the state
is given by (\ref{3}), and entailing a virtual absence of any
discussion of measurement of other observables. This neglect has
played a very confusing role in the discussions following the EPR
proposal. Occasional references to a broader point of view, like
the one by Hooker (\cite{Hooker72}, section~5), to the effect that
consideration of a joint measurement of position of particle $1$
and momentum of particle $2$ might provide an argument to be used
in the EPR challenge of the Copenhagen interpretation, remained
largely unnoticed.

\section{The Einstein-Podolsky-Rosen experiment} \label{sec2.4}
The Einstein-Podolsky-Rosen experiment \cite{EPR} can be
considered as an ultimate attempt from Einstein's part to prove
`incompleteness of quantum mechanics', in the sense that quantum
mechanics is not capable to account for the sharp values of both
position and momentum that allegedly can be simultaneously
attributed to a microscopic object. Earlier attempts had been
defeated by Bohr by pointing to the disturbing influence of
measurement, making such a simultaneous attribution impossible.
Therefore the EPR proposal was devised so as to make it possible
to obtain knowledge about a particle ``without in any way
disturbing the system.'' In the experiment a two-particle system
is considered, described by an entangled state of the type
(\ref{3}), the two particles being prepared so as to be so far
apart that a measurement on particle $1$ does not influence
particle $2$. Due to the correlation between observables
$A^{(s)}_1$ and $B^{(s)}_2$ expressed by (\ref{3}), it is possible
according to EPR to infer the value of the quantum mechanical
observable $B^{(s)}_2$ of particle $2$ from the measurement result
obtained by measuring observable $A^{(s)}_1$ on particle $1$. In
this way information on particle $2$ is obtained without in any
way disturbing this particle.

A conclusion of `incompleteness of quantum mechanics' is drawn by
EPR for the special situation in which the expansion (\ref{3}) of
state vector $| \psi_{12} \rangle$ is not unique, in the sense
that orthonormal sets $\{|\alpha'_{1i}\rangle\}$ and
$\{|\beta'_{2i}\rangle\}$ of eigenvectors of observables $A'_1$
and $B'_2$, respectively, exist such that $| \psi_{12} \rangle$
can be also expressed according to
\[| \psi_{12} \rangle = \sum_i c'_i | \alpha'_{1i} \rangle |
\beta'_{2i} \rangle.\] The crucial point of the EPR reasoning is
that observable $A'_1$ is incompatible with $A^{(s)}_1$ (and
analogously for observables $B^{(s)}_2$ and $B'_2$ of particle
$2$). Because particle $2$ cannot experience which of the two
observables is measured on particle $1$, it is concluded by EPR
that values of the incompatible observables $B^{(s)}_2$ and $B'_2$
can be simultaneously attributed to particle $2$. `Incompleteness
of quantum mechanics' follows from the circumstance that quantum
mechanics is not able to describe a state of particle $2$ in which
both of these observables have sharp values.

It is important to note here that in reaching this conclusion EPR
in an essential way make use of an {\em objectivistic}-realist
interpretation of the mathematical formalism of quantum mechanics,
since they assume quantum mechanical measurement results to
correspond to {\em objective} properties of the microscopic object
(`elements of physical reality'). It is precisely on this count
that Bohr \cite{Bohr35} challenged the EPR conclusion of
incompleteness. According to Bohr's correspondence principle, to
be discussed in section~\ref{sec2.5}, quantum mechanical
measurement results have a {\em contextual} meaning that is
determined only by taking into account the `whole experimental
arrangement', including the measurement arrangement for particle
$1$. The actual presence of this arrangement is essential for
obtaining knowledge on particle $2$. According to Bohr, Einstein's
definition of elements of physical reality of particle $2$ is
ambiguous because it does not take into account the experimental
arrangement of particle $1$. Bohr thought that if this is
corrected the EPR challenge could be dealt with in precisely the
same way earlier attempts by Einstein to prove incompleteness of
quantum mechanics were countered.

However, Bohr's above-mentioned judgment is not completely
reliable (e.g. Popper (\cite{Pop}, p.~149), Jammer (\cite{Jammer},
p.~194), also Folse \cite{Folse85}). Admittedly, the issue of
`contextuality' is used by him also here, but {\em not} in the
{\em interactional} sense involved in the `quantum postulate'.
Bohr's answer to EPR implies a change of interpretation, to the
effect that for defining a physical quantity of particle $2$ it is
sufficient that a {\em relation} exist between particle $2$ and
the measuring instrument (which interacts only with particle $1$).
It is no longer deemed necessary that this relation (which was
compared to a coordinate system by Bohr) be realized by an {\em
interaction}, like it was before.

This change of interpretation from an interactional to a {\em
relational} point of view has had a large impact on later
developments, because it introduced into the interpretation of
quantum mechanics an issue of `nonlocality' which since that time
has stayed with us. It was contended by Einstein that this
nonlocality is just an artefact of Bohr's interpretation,
combining contextuality of observables with an individual-particle
interpretation of the state vector (i.e. completeness). According
to him objectivity of observables could be restored by assuming an
ensemble interpretation of the state vector (i.e. incompleteness):
such an interpretation would allow to avoid the Copenhagen
indeterminism (implying that a value of a quantum mechanical
observable cannot be attributed to the object as a property
possessed prior to measurement), and to interpret von Neumann
projection, applied to particle $2$ on the basis of a measurement
on particle $1$, as a selection of a subensemble. Unfortunately,
in 1935 no Kochen-Specker or Bell theorems were available to put
this into doubt. Nowadays we are convinced by these theorems that
it is not possible to attribute values of all quantum mechanical
observables as objective properties to the object, possessed
independently of measurement, and that, therefore, Einstein's
ensemble interpretation cannot be the final solution (e.g. Guy and
Deltete \cite{GuyDel90}). This does not imply, however, that
Bohr's nonlocal contextuality must be accepted. Its general
acceptance is based on a number of confusions and inconsistencies
which have entered the Copenhagen interpretation, to be discussed
in the following sections.

\section{Correspondence and complementarity} \label{sec2.5}
According to Petersen \cite{Pet}, the correspondence principle is
characteristic of the ``Copenhagen spirit of quantum theory.'' The
idea of `correspondence' is an expression of the necessity felt to
maintain as much as possible the use of classical concepts within
the atomic domain. It is necessary to distinguish a weak and a
strong form of the correspondence principle. In its weak form
(e.g. Messiah \cite{Messiah}, section 1.12) it is a requirement
that the laws of microscopic physics must be formulated in such a
way that in the limit of large quantum numbers (the `classical
limit') quantum mechanical results must agree with the results of
classical mechanics. In its strong form the correspondence
principle is valid not only in an asymptotic sense, but it is a
requirement to be met by {\em any} measurement within the
microscopic domain, independently of the values of the quantum
numbers.

By the strong form of the correspondence principle the necessity
is expressed of basing the description of the properties and
manipulation of the measuring instruments on purely classical
ideas, which are the ideas of macrophysics, thought to be the
only means of unambiguous communication. According to Bohr a
quantum phenomenon can be communicated only by using the {\em
classical} terms by which the measurement arrangement is
characterized. The measurement arrangement plays a
key role in defining a quantum mechanical observable.

\noindent {\em A quantum mechanical observable is exclusively
defined by and within the context of the measurement serving to
measure that observable; experimental arrangement and measurement
results are to be described in classical terms} (strong form of
the correspondence principle).

\noindent Note that according to Bohr this does not mean that a
quantum mechanical measurement process could not be described
quantum mechanically. However, such a description would not be
able to account for the macroscopic behavior of the measuring
instruments, and, for this reason, would no longer allow to
consider the process as a {\em measurement} process.

The idea of `complementarity' is just an extension of the idea of
`correspondence'. If the definition of a quantum mechanical
observable is restricted by its measurement arrangement, then
mutual exclusiveness of measurement arrangements can explain why
incompatible observables cannot simultaneously have sharp values.
Thus,

\noindent {\em Incompatible quantum mechanical observables
correspond to mutually exclusive measurement arrangements,
defining different aspects of reality which cannot be united in a
single classical picture} (complementarity).

\noindent Bohr and Heisenberg applied the idea of
`complementarity' in the first place to position and momentum
observables, which, due to complementarity, cannot be defined more
accurately than is allowed by the Heisenberg uncertainty relations
\begin{equation}\label{7}
\Delta Q\Delta P\geq \hbar/2,\;(\Delta Q)^2= \langle
\Psi|(Q-\langle Q\rangle)^2|\Psi\rangle,\;(\Delta P)^2= \langle
\Psi|(P-\langle P\rangle)^2|\Psi\rangle,
\end{equation}
thus allegedly implementing the idea of `completeness in a
restricted sense' in the quantum mechanical formalism as a
consequence of incompatibility of the corresponding operators:
$[Q,P]_-=i\hbar$. By Heisenberg this was interpreted as a
consequence of mutual disturbance in a simultaneous measurement of
position and momentum, caused by mutual exclusiveness of the
measurement arrangements serving to measure each of these
observables separately in an ideal way.

It should be noticed that the interpretations of Bohr's and
Heisenberg's are rather different from each other. By Heisenberg
the quantities $\Delta Q$ and $\Delta P$ are interpreted as
measures of the measurement accuracies of simultaneously performed
measurements of position and momentum, respectively. Heisenberg
(\cite{Heis30}, section II.2) explicitly noted that inequality
(\ref{7}) must be seen as a property of the microscopic object,
valid in its {\em post}-measurement state, and, therefore, to be
seen as an objective property of the microscopic object {\em
after} it has ceased to interact with the measuring instrument.
For Bohr the inequality marks a limitation of our possibility to
apply to a microscopic object the classical notions of position
and momentum (latitudes of definition) {\em during} a simultaneous
measurement of these quantities. Of course, if every initial state
were prepared by a simultaneous measurement of position and
momentum, the difference between Bohr and Heisenberg would be
inconsequential. However, although all measurements are also
preparations, the converse is not very plausible. Equating
arbitrary preparations realized in nature with measurements would
stretch the definition of `measurement' rather too much. The
seeming agreement on the meaning attributed by Bohr and Heisenberg
to the Heisenberg uncertainty relations, marks the fundamental
failure of the Copenhagen interpretation to distinguish
`preparation' and `measurement' (cf. section~\ref{sec2.7}). It was
realized only a very long time after the inception of the idea of
`complementarity' (Ballentine \cite{Bal70}) that it is
inconsistent to interpret (\ref{7}) -which is a property of the
initial ({\em pre}-measurement) state $|\Psi\rangle$- as referring
to a subsequent measurement.

Although presumably not completely justified, the correspondence
principle has often been considered as demonstrating an
inclination of the Copenhagen interpretation towards
positivism/empiricism. Indeed, the insight that an account of a
quantum mechanical measurement must be based on a so-called
`observation language', or, at least, on an independently tested
theory (viz. classical mechanics), was consistent with the logical
positivist/empiricist ideas. We should be cautious with such a
conclusion, however. Empiricism does not seem to be involved in
Bohr's characterization of the quantum phenomenon by `the
experiment as a whole', supposed to refer to observations obtained
under specific circumstances including an account of the whole
experimental arrangement. Nor does the Copenhagen usage of
interpreting measurement results as (classical) properties of the
microscopic object rather than as pointer readings of a
(macroscopic) measuring instrument point into the direction of
empiricism. Above all, the non-vanishing value of the `quantum of
action' is the characterizing feature stressed by Bohr, making it
impossible to draw a sharp distinction between microscopic object
and measuring instrument. Heisenberg (\cite{Heis58c}, p.~145),
while distinguishing between the reality of the positivistic sense
impressions of an observer and the reality of objects and events
dealt with in atomic physics, even explicitly denies that the
Copenhagen interpretation would be a positivistic one (see also
section~\ref{sec3}).

Nowadays the reliance of the correspondence principle on a {\em
classical} description of measurement is recognized as a severe
drawback. It has been realized that it is impossible to describe
information transfer from a microscopic object to a measuring
instrument by means of classical mechanics alone. Admittedly, a
measuring instrument for measuring a quantum mechanical observable
must have a macroscopic pointer, to be described classically as
far as its macroscopic properties are concerned. But it is equally
important that a measuring instrument have also a {\em
microscopic} component which is sensitive to the microscopic
information that in the measurement process is transferred to it
from the microscopic object. Actually, quantum mechanics presents
a clear paradigm of the `theory-ladenness of observation', which
during the second half of the $20^{th}$ century caused the decline
of logical positivist/empiricist influence. A growing awareness of
the necessity to consider a physical theory not as universally
valid, but only as valid on a certain domain of application, has
stimulated the view that application of classical notions to
microscopic processes may not be such a good idea in general. The
information transfer between the microscopic object and the
sensitive part of the measuring instrument is a microscopic
process, well within the domain of application of quantum
mechanics.

It should be noted that the ban on a quantum mechanical
description of measurement, imposed by the correspondence
principle, has been disregarded e.g. by von Neumann and by
Heisenberg. However, the quantum mechanical treatments by von
Neumann and Heisenberg probably had the intention to justify the
idea of `correspondence' by means of quantum mechanical
considerations, rather than to be independent attempts at a
quantum mechanical account of measurement. As will be reviewed in
section~\ref{sec2.8}, a genuine application of quantum mechanics
to quantum mechanical measurement is indispensable for an
assessment of the role played by measurement in quantum mechanics.
In particular, it follows that a generalization of the
mathematical formalism is necessary in order to be able to
describe even experiments that -like the two-slit experiment- were
crucial in developing the idea of `complementarity'. Also, von
Neumann's projection postulate turns out to be obsolete as a
measurement principle securing a well-defined value of an
observable as a result of a measurement. It seems that the urge to
interpret such a result as a property of the microscopic object is
a consequence of the classical thinking involved in the
correspondence principle.

The correspondence principle is consistent with the idea that a
quantum mechanical observable does not have a value prior to its
measurement. The contextual meaning of a quantum mechanical
observable is justified by the essential role played by the
measurement arrangement in its definition (cf.
section~\ref{sec2.1.2}). It seems that theorems like the
Kochen-Specker and Bell theorems corroborate this idea. Therefore,
in our attempt to develop a neo-Copenhagen interpretation of the
quantum mechanical formalism this feature will be maintained:
quantum mechanical observables are associated with the measurement
arrangements set up to realize their measurement. Actually, since
this feature is perhaps the prime characteristic of the Copenhagen
interpretation, this is a reason to refer to the new
interpretation as a neo-Copenhagen one, even though the latter
will differ from it in many respects. In particular, I will take
seriously the empiricist connotation of the correspondence
principle mentioned above, by associating quantum mechanical
observables with properties of the measuring instrument rather
than with (contextual) properties of the microscopic object. This
makes the essential role of the measurement arrangement in
interpreting the quantum mechanical formalism even stronger. By
doing so it will be seen in section~\ref{sec2.8} that the idea of
`complementarity' as a consequence of mutual disturbance in a
joint measurement of incompatible observables need not be proposed
as a separate principle, but that it straightforwardly follows
from the mathematical formalism of quantum mechanics.

\section{EPR and correspondence}\label{sec2.5.1}
There is an important lesson to be learnt from Bohr's
\cite{Bohr35} application of the correspondence principle in his
answer to the EPR challenge. Actually, this application does not
seem to be in complete agreement with this principle as given
above. By Einstein it was assumed that in the state (\ref{3})
observables $A_1^{(s)} $ and $B_2^{(s)} $ are strictly correlated
so as to allow inference of the value of one observable from the
result obtained by measuring the other one. It is unfortunate that
Bohr did not challenge this assumption on the basis of a strict
application of the strong form of his correspondence principle. On
the basis of this principle Bohr could have rejected the
possibility of defining the {\em correlation} of observables
$A_1^{(s)} $ and $B_2^{(s)} $ outside the measurement context for
measuring it. A strict application of the correspondence principle
(strong form) would have implied the requirement that measurements
be performed on {\em both} particles, thus making obsolete the EPR
idea of applying the notion of an `element of physical reality',
and, consequently, blocking the whole EPR reasoning. By taking
into account that a definition of the correlation would imply that
each of the particles is interacting with its own measuring
instrument, Bohr could have maintained his contextualistic-realist
interpretation in the {\em interactionist} sense originally
intended, without any need to recede to a relational point of view
(cf. section~\ref{sec2.4}). Moreover, there would not have been
any reason to resort to nonlocality in order to implement the
contextuality going with the ``nonlocal'' measurement arrangement,
since each particle's context can be {\em locally} determined by
the measuring instrument the particle is interacting with.

However, Bohr did not challenge EPR in this way. Evidently, he did
not recognize the {\em correlation} of observables $A_1^{(s)} $
and $B_2^{(s)} $ as an ordinary observable, to be measured by
means of coincidence measurements like the EPR-Bell ones performed
e.g. by Aspect et al. \cite{Asp81,Asp82}. Instead, he followed EPR in
taking for granted existence of the correlation of the observables
already if only one of the observables is measured. Consequently,
in order to apply his correspondence principle Bohr had to
refer to the context furnished for particle $2$ by the measurement
arrangement of particle $1$, thus starting the nonlocality enigma.
On a consistent application of the correspondence principle we
would not have had any reason to infer any nonlocality if
measurements of the EPR-Bell type are considered rather than
experiments of the EPR type.

Of course, this does not imply that such an inference of
nonlocality could not be drawn in a different way from the EPR
experiment as it was actually proposed. As already remarked in
section~\ref{sec2.4}, it is not possible to accept EPR's proposal
to solve all problems by means of an ensemble interpretation of
the state vector. In order to analyze this, it is necessary,
however, to discuss another confusion inherent in the Copenhagen
interpretation, viz. the confusion of the notions of `preparation'
and `measurement', which also Bohr has seemingly fallen prey to in
analyzing the EPR proposal, by accepting it as a measurement of an
observable of particle $2$ rather than as a preparation applied to
this particle.

\section{Preparation and measurement} \label{sec2.7}
Since Bohr was ready to apply his correspondence principle (strong
form) to the EPR experiment, he evidently accepted it as a {\em
measurement} of a property of particle $2$, in the way intended by
EPR. It is questionable, however, whether this is justified.
Admittedly, on the basis of the existence of the strict
correlation between observables $A_1^{(s)} $ and $B_2^{(s)} $, as
suggested by (\ref{3}), the measurement result of observable
$B_2^{(s)} $, to be obtained if it were actually carried out in
coincidence with the measurement of $A_1^{(s)} $, could be
inferred already from the result of the measurement of the
particle $1$ observable. However, this inference is possible only
if the correlation is assumed to exist {\em independently of its
measurement}. Only on this basis can the measurement of particle
$1$ be interpreted as a measurement of the correlated property of
particle $2$. To the extent the strong form of the correspondence
principle is an essential ingredient of the Copenhagen
interpretation, this entails an internal inconsistency of this
interpretation if it accepts the EPR experiment as a {\em
measurement} of particle $2$.

In order to remove this inconsistency, it is necessary to duly
distinguish between the EPR experiment and EPR-Bell experiments
like the ones performed by Aspect et al., in the sense that in the
former there is {\em no measurement} carried out on particle $2$.
Instead, by the EPR procedure this latter particle (better: the
corresponding ensemble) is {\em prepared} in a certain state,
conditional on a measurement result yielded by the measurement of
particle $1$. Not distinguishing between the processes of
`measurement' and `conditional preparation' has caused quite a bit
of confusion. For instance, in an EPR-Bell experiment it would be
possible to measure in coincidence with $A_1^{(s)} $ an observable
$B_2'$ of particle $2$, incompatible with the special observable
$B_2^{(s)} $. If the measurement of $A_1^{(s)} $ were still to be
interpreted as a measurement of $B_2^{(s)} $, this would imply a
simultaneous measurement of two incompatible observables of
particle $2$, which is in disagreement with another assumption of
the Copenhagen interpretation, viz. the complementarity principle.
As this latter principle is a cornerstone of the Copenhagen
interpretation which is to survive (cf. section~\ref{sec2.8}), it
seems wise to prevent such a disagreement by duly distinguishing
EPR experiments from EPR-Bell ones.

Unfortunately, the confusion of the notions of `preparation' and
`measurement', observed here, is a general feature of the
Copenhagen interpretation, induced by the impossibility of
attributing an objective value of a quantum mechanical observable
to the microscopic object, possessed prior to measurement. As a
consequence, a measurement cannot reveal a pre-existing value of
an observable, as is generally thought to be the case in classical
mechanics. As a second best solution to this problem it was
assumed that, at least, the value of the observable may be
attributed to the object {\em immediately after the measurement}.
This, actually, is the origin of von Neumann's projection
postulate, intending the measurement to {\em prepare} the object
in an eigenstate of the measured observable. Application of this
postulate to the EPR experiment, considered as a measurement of
observable $B_2^{(s)} $, has been responsible for the idea that
the particle $2$ state is projected onto one of the eigenstates
$|\beta^{(s)}_{2i}\rangle$. This idea has already been criticized
by Margenau \cite{Marg36} on the basis that von Neumann's
projection postulate is necessary only in an individual-particle
interpretation of the wave function, and would be meaningless in
an ensemble interpretation because a measurement on an individual
particle cannot change the state of a whole ensemble.

Notwithstanding Margenau's criticism, the idea of a quantum
mechanical measurement as preparing the microscopic object in one
of a number of macroscopically distinguishable states, thus
allowing to obtain a measurement result by ascertaining `in which
of these states the object finally is' (like, for instance, the
Stern-Gerlach experiment), has become more or less paradigmatic of
the Copenhagen interpretation. For instance, in his axiomatization
of quantum mechanics, Jauch (\cite{Jauch}, section~11-3) considers
a measurement as a {\em filter}, leaving the object in one of the
eigenstates of the measured observable (so-called measurements of
the first kind). By now it is well known, however, that the large
majority of measurement procedures employed in actual practice
does not satisfy this model. In general, it is not the microscopic
object itself, but rather the pointer of a measuring instrument
that is brought into one of a set of macroscopically
distinguishable states. The microscopic object may even be
annihilated in the measurement process (like, for instance,
photons detected by an ideal photon counter). Conditional on the
final pointer state the object is left behind in some state
determined by the interaction of object and measuring instrument,
in general differing from an eigenvector of the measured
observable. For this reason it is hardly appropriate to consider
von Neumann's projection postulate as a {\em measurement}
principle, generally valid in quantum mechanical measurement.

However, this reasoning does not apply to EPR, because this
experiment is not a {\em measurement} of particle $2$. It is
precisely due to this fact that von Neumann's projection postulate
can be applied to EPR. Indeed, this postulate is a valid {\em
preparation} principle, routinely applied in the laboratory for
preparing microscopic objects in well-defined quantum mechanical
states. Thus, if in the two-particle state (\ref{3}) a measurement
of an arbitrary observable $F_2$ of particle $2$ is carried out,
conditional on measurement result $a_{1i}$ of the jointly measured
observable $A_1^{(s)} $, then the conditional probability
$p(f_{2j}|a_{1i})$ of measurement result $f_{2j}$ of $F_2$ can be
derived from the joint probability
\begin{equation} p(a_{1i},f_{2j}) = |\langle\psi_{12}|
\alpha^{(s)}_{1i}\phi_{2j}\rangle|^2\label{4}
\end{equation}($|\phi_{2j}\rangle$ the eigenvector of $F_2$
corresponding to eigenvalue $f_{2j}$), according to
\begin{equation} p(f_{2j}|a_{1i}) =
\frac{p(a_{1i},f_{2j})}{p(a_{1i})}.\label{5}
\end{equation}
Putting
\begin{equation} p(f_{2j}|a_{1i})= \langle\phi_{2j}| \rho_{2i}
|\phi_{2j}\rangle,\label{6}
\end{equation}
it is straightforwardly proven that the density operator
$\rho_{2i}$ of the particle $2$ state satisfying (\ref{6}) is
given by $|\beta^{(s)}_{2i}\rangle\langle \beta^{(s)}_{2i}| $.
Hence, the subensemble of particles $2$ prepared by selecting
these particles on the basis of measurement result $a_{1i}$ of
observable $A_1^{(s)} $ of particle $1$ from an ensemble of
particle pairs prepared in state (\ref{3}), is described by the
state vector $|\beta^{(s)}_{2i}\rangle$ postulated by von
Neumann's projection postulate. Evidently, as a principle of
conditional {\em preparation} this postulate has a certain
legitimacy. It should be stressed that this legitimacy derives
from the fact that the final state of the preparation process of
particle $2$ is not influenced by any interaction with a measuring
instrument.

Measurements of the EPR type are sometimes called `predictive
measurements' (e.g. Kemble (\cite{Kemble}) to distinguish them
from `determinative measurements' yielding only information about
probabilities of the initial state, without telling anything about
the post-measurement state of the microscopic object. In the
former it is precisely this latter state that is the important
issue, in the sense that we have here a method to prepare,
conditional on the measurement result, the microscopic object in a
well-defined state (like $|\beta^{(s)}_{2i}\rangle$ in the EPR
procedure). Here the {\em preparation} is the crucial issue: in
the post-measurement state a (subsequent) measurement of an {\em
arbitrary} observable $F_2$ can be performed. The EPR experiment
is a (conditional) preparation of particle $2$. Its interpretation
as a measurement of an observable of this latter particle has
caused much confusion. In particular, it seduced Bohr to apply his
correspondence principle (strong form), even though it is a {\em
measurement} principle rather than a {\em preparation} principle.
Evidently, Bohr did not consider the observable $A_1^{(s)} \otimes
B_2^{(s)} $, corresponding to a measurement of the correlation of
the observables of particles $1$ and $2$, as an ordinary quantum
mechanical observable, to be well-defined only within the context
of the experimental arrangement for measuring it. By interpreting
the conditional preparation of particle $2$ as a measurement he
went even so far as to assume that the correlation between the
particles is determined by the measurement arrangement for only
one of the particles (as well as, of course, by the preparing
apparatus). The nonlocal contextuality ensuing from this must be
seen as a consequence of an inconsistent application of the
correspondence principle (strong form).

\section{Conditional preparation versus contextual state}
\label{sec2.7.1} The fundamental confusion of preparation
and measurement, involved in the Copenhagen interpretation, can
be highlighted in yet another way. For this purpose let us define
for a system, prepared with state $\rho$, and on which a
measurement is performed of standard observable $A=\sum_m a_m
P_m$, the {\em contextual state}
\begin{equation}\label{2.7.1.1}
\rho_A=\sum_m P_m \rho P_m.
\end{equation}
It is usual to consider this state as the final state of a von
Neumann-L\"uders measurement procedure, for which, in agreement
with Heisenberg's ideas discussed above, the equality $Tr \rho_A
P_m= Tr \rho P_m$ allows to infer the probabilities of the initial
state from those measured in the final one. However, in view of
the inapplicability of the notion of `measurement of the first
kind', this interpretation does not seem to be very useful. Bohr's correspondence principle (strong form) offers a way to look upon the state
$\rho_A$ (\ref{2.7.1.1}) in a different way, viz. as an
alternative description of the initial state of an object,
prepared according to $\rho$, as soon as it is within the context
of the measurement arrangement for measuring observable $A$. Note
that, due to Bohr's instrumentalist conception of quantum
mechanical state vectors, presumably this interpretation is far
from Bohrian. But it would at least yield another ontological
implementation of Bohr's idea that observable $A$ is well-defined
only within the context of a measurement of this observable,
consistent with Bohr's correspondence view of observables
discussed in section~\ref{sec2.5}. It would allow within quantum
mechanics to distinguish between the state of an object as it is
prepared prior to its interaction with a measuring instrument, and
the state probed by this measuring instrument, possibly already
influenced by that instrument. This is quite analogous to the way
a classical rigid body model of a billiard ball accounts for
certain observations, thus accounting for rigidity as a property
of a billiard ball {\em only to be attributed within a particular
experimental context} (cf. section~\ref{sec2.1.2}; see
section~\ref{sec4} for another analogy).

The contextual state should be distinguished from the
conditionally prepared states of a measurement procedure. The
distinction between a contextual state and a conditionally
prepared state is particularly telling in the EPR experiment. As
demonstrated in section~\ref{sec2.7}, in a measurement of the
particle $1$ observable $A_1^{(s)} $ in the two-particle state
(\ref{3}), particle $2$ is conditionally prepared in the state
$|\beta^{(s)}_{2i}\rangle$. In the Copenhagen interpretation this
is often interpreted in the sense that the corresponding value of
observable $B_2^{(s)} $ has got a definite value due to the
measurement on particle $1$. In Bohr's correspondence view of the
EPR experiment it is even assumed that this value must be
well-defined as soon as particle $1$ is interacting with its
measuring instrument, thus starting the nonlocality enigma.

Now the question can be asked: does the state vector
$|\beta^{(s)}_{2i}\rangle$ ``really'' describe the reality of
particle $2$ as it is in the context of a measurement of an
observable of particle $1$? The problem is that we cannot tell. In
order to probe this reality we have to make a measurement on
particle $2$ of some observable $F_2= \sum_j f_{2j}
|\phi_{2j}\rangle\langle \phi_{2j}|$. In the context of this
measurement the contextual state is given by
\[|\beta^{(s)}_{2i}\rangle\langle\beta^{(s)}_{2i}|_{F_2} =\sum_j
|\langle
\beta^{(s)}_{2i}|\phi_{2j}\rangle|^2|\phi_{2j}\rangle\langle
\phi_{2j}|.\] This state is very different from
$|\beta^{(s)}_{2i}\rangle\langle \beta^{(s)}_{2i}|$. Since these
states yield the same measurement results for $F_2$, it is
impossible to tell which is the ``real'' one. Whether before the
measurement the state ``really'' was
$|\beta^{(s)}_{2i}\rangle\langle \beta^{(s)}_{2i}|$ is a matter
of interpretation. This will be discussed more extensively in
section~\ref{sec3}, where a different interpretation will be proposed.

More generally, for the two-particle system involved
in an EPR-Bell measurement in which an arbitrary correlation
observable $A_1\otimes B_2 = \sum_{i}
a_{1i}|\alpha_{1i}\rangle\langle\alpha_{1i}| \otimes \sum_j b_{2j}
|\beta_{2j}\rangle\langle\beta_{2j}| $ is measured in an arbitrary
state $|\psi_{12}\rangle$, the contextual state (\ref{2.7.1.1}) is
given by
\[|\psi_{12}\rangle\langle\psi_{12}|_{A_1\otimes B_2}= \sum_{ij}
|\langle\psi_{12}|\alpha_{1i}\beta_{2j}\rangle|^2
|\alpha_{1i}\beta_{2j}\rangle\langle\alpha_{1i}\beta_{2j}|.\] From
this expression the contextual states of the two particles
separately can be derived by partial tracing as the contextual
states to be found for each particle, independently of which
measurement is carried out on the other particle. This illustrates
the possibility of {\em local} contexts referred to in
section~\ref{sec2.5}. Contrary to what is often supposed, the
conditionally prepared state need not play any prominent role in
describing the reality of particle $2$ if a measurement is carried
out on this particle, because in that case this reality is locally
(co-)determined by the measurement arrangement that is actually
present at the location of this particle.

\section{Generalized observables, and complementarity} \label{sec2.8}
As already noted in section~\ref{sec2.5}, the Copenhagen confusion
of the notions of `preparation' and `measurement' has had an
equally confusing effect with respect to the issue of
`complementarity'. By Heisenberg `mutual disturbance in a joint
measurement of incompatible observables' was taken in a {\em
preparative} sense, in which  `disturbance' is referring to the
preparation of the final state of the object. There is another
possibility, however, to the effect that `disturbance' may refer
to the final state of the {\em measuring instrument}, in the sense
that the probability distribution of the final pointer positions
may depend on the details of the measurement procedure, and,
hence, may change if the measurement arrangement is changed so as
to also yield information on an observable incompatible with the
measured one. Then `measurement disturbance' may refer to a
deviation of such a probability distribution from the ``ideal''
one obtained in a measurement procedure reproducing the
probability distribution predicted by the standard formalism of
quantum mechanics. This will be referred to as `mutual disturbance
in a determinative sense'. Due to a restriction of the attention
to measurements of the Heisenberg type in which measurement
results are associated with final states of the microscopic
object, the distinction between the preparative and determinative
types of disturbance has remained largely unnoticed.

This subject has been more fully discussed elsewhere
\cite{dM2000,dM2002}, and will be reviewed here only briefly, even though
it is of primary importance for understanding the ways in which
the Copenhagen interpretation has been confused. It is important
to note here that a major cause of this confusion was the fact
that at the time Bohr and Heisenberg developed the complementarity
principle the quantum mechanical formalism had not been fully
developed. Conclusions with respect to uncertainties were often
drawn from the {\em classical} reasoning advocated by the
correspondence principle. As far as measurement was described
quantum mechanically at all, this description seems to have served
mainly to justify the idea of `correspondence' (in particular, by
means of von Neumann projection). Only after the {\em quantum
mechanical} character of measurement in the atomic domain has been
taken seriously, it has become possible to straighten out the
above-mentioned confusion involved in the notion of
`complementarity'. It turns out that complementarity in the sense
of `mutual disturbance in a joint measurement of incompatible
observables' need not be seen as a consequence of limitations of
the classical reasoning involved in the correspondence principle,
but that, if `mutual disturbance' is taken in the
determinative sense defined above, it can be derived from the mathematical formalism of quantum mechanics.

It had to be realized first,
however, that the standard representation of a quantum mechanical
observable as a Hermitian operator is too restricted a concept to
encompass all possible measurements in the atomic domain (in
particular, joint measurement of incompatible observables). This
follows straightforwardly from a quantum mechanical treatment of
the interaction of a microscopic object and a measuring
instrument, by applying the usual measurement postulates to the
measuring instrument. We find the quantum mechanical probabilities
according to
\begin{equation}\label{7.1}
    p_m=Tr_o \rho M_m,\; M_m=Tr_a \rho_a U^\dagger E_m^{(a)}U,
\end{equation}
in which $\rho_a$ is the initial state of the measuring
instrument, $U$ is the unitary evolution operator of the
interaction process, and $E_m^{(a)}$ are operators of the
measuring instrument $a$ determining the detection probabilities
$p_m$ of the measurement as their final state expectation values.
Compared to the standard formalism of textbook quantum mechanics
the generalization consists of the fact that nothing requires the
operators $M_m$ to be projection operators. In general, the
quantum mechanical measurement results (detection probabilities)
are represented according to (\ref{7.1}) by the expectation values
of a positive operator-valued measure (POVM) $\{M_m\}$ taken in
the {\em initial} state $\rho$ of the microscopic object. It can
be verified that the two-slit experiment -consideration of which
has contributed in an essential way to the development of the
notion of `complementarity'- is not representable by a Hermitian
operator (compare de Muynck \cite{dM2002}, section~7.3). It is not
surprising that much confusion has been generated by drawing
general conclusions from a too restricted mathematical formalism,
viz. the standard formalism restricting to the projection-valued
measures (PVMs) corresponding to the spectral representations of
Hermitian (better: self-adjoint) operators. In particular it has
led to unjustifiedly interpreting inequality (\ref{7}) as a
consequence of mutual disturbance in a joint measurement of
position and momentum, rather than as a representation of our
restricted ability to {\em prepare} initial states (cf. Ballentine
\cite{Bal70}).

As demonstrated by Martens and de Muynck \cite{MadM90}, the
inapplicability of the Heisenberg inequality (\ref{7}) to mutual
disturbance in a joint measurement of incompatible observables
does not imply that this latter feature of quantum mechanical
measurement is not a perfectly real one. As a matter of fact, this
feature has been demonstrated to exist in a number of `thought
experiments' widely discussed during the days of the inception of
the idea of `complementarity'. However, for its theoretical
description it needs the generalization of the standard formalism
referred to above. Whereas it is not at all clear how a joint
measurement of incompatible standard observables could be
described by the standard formalism, it is easy to define a joint
measurement of incompatible observables (POVMs) $\{P_m\}$ and $\{
Q_n\}$ by requiring that a bivariate POVM $\{R_{mn}\}$ exist, of
which $\{P_m\}$ and $\{ Q_n\}$ are marginals, such that the
expectation value $Tr \rho R_{mn}$ can be interpreted as the joint
probability distribution of the latter observables. It is easy to
find examples satisfying this requirement. In agreement with the
standard formalism, this is impossible if $\{P_m\}$ and $\{ Q_n\}$
are projection-valued measures of standard observables,
represented by the corresponding Hermitian operators.

More generally, it is possible to define a joint {\em nonideal}
measurement of incompatible POVMs $\{P_m\}$ and $\{ Q_n\}$ by
requiring a bivariate POVM $\{R'_{mn}\}$ to exist, such that
\begin{equation}\label{8}
\begin{array}{l}
\sum_{n} R'_{mn} = \sum_{m'} \lambda_{mm'} P_{m'},\; \lambda_{mm'}
\geq 0,\; \sum_{m} \lambda_{mm'} = 1,\\
\sum_{m} R'_{mn} = \sum_{n'} \mu_{nn'} Q_{n'},\; \mu_{nn'} \geq
0,\; \sum_{n} \mu_{nn'} = 1.\end{array}
\end{equation}
Here matrices $(\lambda_{mm'})$ and $(\mu_{nn'})$ are so-called
nonideality matrices, defining the nonideality of the
determination of probabilities of observables $\{P_m\}$ and $\{
Q_n\}$, respectively, if POVM $\{R'_{mn}\}$ is measured. It should
be noted that $\{P_m\}$ and $\{ Q_n\}$ may be PVMs here.

As measures of nonideality of the nonideality matrices
$(\lambda_{mm'})$ and $(\mu_{nn'})$ it is useful to take the
average row entropies (restricting to finite dimension $N$)
\begin{equation}\label{9}
J_{(\lambda)} = - \frac{1}{N} \sum_{mm'} \lambda_{mm'} \ln
\frac{\lambda_{mm'}} {\sum_{m''} \lambda_{mm''}}
\end{equation}
(and analogously for $(\mu_{nn'})$). If $\{P_m\}$ and $\{ Q_n\}$
are PVMs, it is possible \cite{MadM90} to derive for the
nonideality measures $J_{(\lambda)}$ and $J_{(\mu)}$ the Martens
inequality
\begin{equation}\label{10}
  J_{(\lambda)} + J_{(\mu)} \geq -\ln \{\max_{mn} Tr P_m
Q_n\}.
\end{equation}
 Note that the right hand
side of inequality (\ref{10}) vanishes if $\{P_m\}$ and $\{ Q_n\}$
are compatible PVMs.

It should be stressed that the meaning of the Martens inequality
is completely different from that of the Heisenberg inequality
(\ref{7}). In contrast to the latter inequality, the Martens
inequality (\ref{10}) is derived from the properties of the
observables {\em alone}. It is a property of the POVM
$\{R'_{mn}\}$, which is independent of the initial state. The
inequality is in an unambiguous way expressing the notion of
`mutual disturbance of measurement results in a joint measurement
of incompatible observables', to the effect that the quantities
$J_{(\lambda)}$ and $J_{(\mu)}$ can be seen as measures of the
inaccuracies of the measurements of the observables $\{P_m\}$ and
$\{ Q_n\}$, respectively, caused by a mutual disturbance of these
observables by the measurement process. In contrast to the
Heisenberg inequality the Martens inequality is a perfect
mathematical representation of the idea of `complementarity' as
illustrated by the `thought experiments' (like the two-slit
experiment and Heisenberg's $\gamma$-microscope). That this role
was initially attributed to the Heisenberg inequality must be seen
as an instance of `jumping to conclusions', in the sense that the
availability of the Heisenberg inequality seemed to corroborate so
perfectly the idea of `complementarity' as defined above that it
clouded most physicists critical senses. Apart from a few critics
like Einstein and Margenau, it had to wait for Ballentine's 1970
paper for an in-depth analysis of the inappropriateness of this
attribution, by demonstrating that the Heisenberg inequality
(\ref{7}) is not a property of a `simultaneous or joint {\em
measurement} of incompatible observables' but rather a property of
{\em `preparation'} (i.e. of the initial state $|\Psi\rangle$).

That the confusion could last so long, has several causes. First
of all, the Martens inequality (\ref{10}) needs the generalized
formalism for its derivation. This formalism was not available
then. Moreover, the idea of `completeness in a wider sense',
applied to the standard formalism of quantum mechanics, may have
been responsible for the acceptance of the Heisenberg inequality
as the only candidate for representing so obvious a physical
phenomenon as `measurement inaccuracy due to mutual disturbance'.
Only after the generalized formalism had been developed it was
possible to break away from this paradigm by deriving from the
generalized formalism the Martens inequality (\ref{10}) as a
better candidate for this purpose. In certain of the `thought
experiments' (for instance, the two-slit experiment with a moving
screen) the measurement inaccuracy is a consequence of the
uncertainties of position and momentum in the initial state of the
{\em screen}, expressed by the Heisenberg uncertainty relation for
the initial state of the screen. This makes it understandable that
measurement inaccuracy was associated with the Heisenberg
inequality. Another reason may be the confusion of preparation and
measurement inherent in a restriction of the attention to
measurements of the first kind, for which it is not implausible to
interpret -as was done by Heisenberg- the standard deviations of
the {\em final} state as measures of inaccuracy of measurement
results to be attributed to the initial state.

Such confusions could arise because it was thought that no quantum
mechanical analysis of the measurement process could be given (cf.
section~\ref{sec2.5}). It must be stressed that the developments
leading to the generalized formalism of quantum mechanics could
take place only on the basis of a rejection of this tenet of the
Copenhagen interpretation. Only on this basis it was possible (de
Muynck \cite{dM2000}) to realize that not one but two
complementarity principles exist, one for preparation (expressed
by the Heisenberg inequality), and one for the joint measurement
of incompatible observables (expressed by the Martens inequality),
both derivable from the (generalized) formalism.

\section{Realist versus empiricist interpretation of quantum
mechanics}\label{sec3} An interpretation of a physical theory is a
mapping from its mathematical formalism into the physical world.
Roughly speaking there are two different possibilities for quantum
mechanics, viz. a realist interpretation and an empiricist one.
Here the possibility is ignored of an instrumentalist
interpretation, to the effect that no mapping into reality is
required at all but the theory is considered as `an instrument for
generating measurement results', because in the author's view this
interpretation's omission of attributing a well-defined physical
meaning to quantum mechanical concepts has been a source of
confusion. This holds true in particular with respect to the
notion of a `quantum mechanical measurement result' which, due to
instrumentalist vagueness, could be taken at will either as a
property of the microscopic object, or as a property of the
measuring instrument, or both.

\noindent {\em In a realist interpretation of quantum mechanics
the mapping is thought to be from the mathematical formalism into
{\em microscopic} reality. The mathematical entities of the theory
(state vector $|\Psi\rangle$, density operator $\rho$, standard
observable $A$, and generalized observable $\{M_m\}$) are thought
to represent properties of the microscopic object} (realist
interpretation of quantum mechanics).

\noindent A realist interpretation is very similar to the way
classical mechanics is usually interpreted. It is the
interpretation adopted in virtually all textbooks of quantum
mechanics. Instruments used to prepare the microscopic object in
an initial state, as well as measuring instruments, are not
represented in the quantum mechanical description even if they are
physically present.

As discussed already in section~\ref{sec2.1}, we should
distinguish objectivistic and contextualistic versions of a
realist interpretation of quantum mechanics. In an
objectivistic-realist interpretation the quantum mechanical
description is thought to refer to objective reality, that is, a
reality independent of any observer, including his measuring
instruments. In a contextualistic-realist interpretation quantum
mechanical concepts are thought to have a meaning only within a
certain physical context (like the object's environment, or a
measurement arrangement). Presumably as a consequence of an
ill-understood classical paradigm, textbook presentations of
quantum mechanics are generally cast into an objectivistic-realist
terminology. However, as demonstrated by the billiard ball analogy
discussed in section~\ref{sec2.1}, not even classical theories do
allow an objectivistic-realist interpretation.

\noindent {\em In an empiricist interpretation of quantum
mechanics the mapping of $|\Psi\rangle, \rho, A$, and $\{M_m\}$
is thought to be from the mathematical formalism into the {\em
macroscopic} reality of instruments or procedures for preparing
and measuring microscopic objects} (empiricist interpretation of
quantum mechanics).

\noindent Thus, in an empiricist interpretation a wave function or
a density operator is thought to be a symbolic representation of a
preparing apparatus (for instance, a cyclotron with specified knob
settings) or a preparation procedure. A quantum mechanical
observable (POVM or PVM) is thought to be a label of a measuring
instrument (for instance, a photosensitive device for detecting
photons) or a measurement procedure.

\noindent Even though the microscopic object is present, in an
empiricist interpretation it is not thought to be represented by
the quantum mechanical formalism. The quantum mechanical formalism
is thought to describe `just the phenomena', phenomena being
situated in the macroscopic instruments for preparing and
measuring the microscopic object. The empiricist interpretation as
defined above is the weakest interpretation of quantum mechanics
deploying a well-defined mapping of the mathematical formalism
into reality. It comes closest to the way quantum mechanics is
being used in the actual practice of experimental physics, if the
physicist's tendency to devise (classical) pictures behind the
phenomena is waived. The quantum mechanical formalism is thought
not to provide any (causal) mechanisms explaining why a certain
measurement result is obtained on a certain individual
preparation. In particular, the formalism is thought to just {\em
describe} EPR correlations, without expecting quantum mechanics to
yield any causal explanation. Note that by itself this does not
imply indeterminism as discussed in section~\ref{sec2.1.3}.
Quantum mechanics is thought to be neutral with respect to this
latter issue, simply because the concept of a quantum mechanical
observable does not apply to preparation procedures.

Notwithstanding similarities, an empiricist interpretation of
quantum mechanics should be carefully distinguished from the
empiricist view as fostered by the philosophical doctrine of
logical positivism/empiricism, which considers metaphysical
anything that is unobservable. In particular, an empiricist
interpretation of quantum mechanics is perfectly consistent with a
belief in a ``real'' existence of microscopic objects like atoms
and electrons (reality behind the phenomena), even if these are
not directly observed, nor described by this theory. Hence, an
empiricist interpretation of quantum mechanics is not to be
confused with the anti-realist philosophy referred to in
section~\ref{sec2.1.1}. According to an empiricist interpretation
quantum mechanics does not describe the microscopic objects
themselves, but just relations between preparing and measuring
procedures, mediated by microscopic objects. It need not be
assumed that correlations like those in EPR-Bell experiments do
not have causes (as is done e.g. by van Fraassen \cite{vF91}).
Quantum mechanics just does not describe these causes. This can be
compared to Newton's interpretation of his theory of gravitation,
to the effect that this theory describes `just the phenomena', and
does not yield any (mechanical) explanation of the way the
gravitational force is transmitted from the sun to a planet (about
which Newton did not wish ``to frame hypotheses''). The
`nonlocality' problem of quantum mechanics, started by the EPR
challenge, is to be compared to the `actio-in-distans' problem of
the Newtonian theory of gravitation: both are consequences of a
realist interpretation of the theory, asking from the theory
explanations it is not able to give.

Hence, an empiricist interpretation of quantum mechanics is not at
variance with metaphysical realism (``the theory that the objects
of scientific enquiry exist and act, for the most part, quite
independently of scientists and their activity'' (Bashkar
\cite{Bashkar})). However, in order to describe the microscopic
objects themselves new (subquantum) theories have to be developed,
analogous to the field theories that
nowadays are thought to be able to yield an explanation of
Newton's `action-at-a-distance'. It should be stressed that, far
from denying the possibility of subquantum (hidden-variables)
theories, an empiricist interpretation of quantum mechanics leaves
considerably more room for such theories than is allowed for by a
realist interpretation, precisely because quantum mechanics is not
thought to describe microscopic reality as such.

The empiricist interpretation should also be distinguished from
the Copenhagen interpretation, which, although having an
empiricist reputation, actually contains many realist elements.
Nevertheless, the empiricist interpretation is indebted to the
Copenhagen one by taking seriously the importance attributed to
the role of the measuring instrument in assessing the meaning of
the quantum mechanical formalism. But the empiricist
interpretation differs from the Copenhagen one by not
considering a quantum mechanical measurement result to be a
property of the {\em microscopic} object (either before, during, or
after the measurement), but to correspond to a {\em macroscopic}
event in the macroscopic part of a measuring instrument (pointer)
that can be recorded in an unequivocal way either by direct
observation or by means of registration by some memory device.
Hence, the information transfer from the microscopic object to the
measuring instrument must be taken into account. In agreement with
the criticism of the correspondence principle given in
section~\ref{sec2.5}, the quantum mechanical formalism is applied
to the interaction of object and measuring instrument.

One reason not to equate the professed empiricism of the
Copenhagen interpretation with an empiricist interpretation of
quantum mechanics as defined here, is that the discussion between
Bohr and Einstein took place completely within the confines of a
realist interpretation. For Einstein this was the natural
interpretation, in which the reality described by a wave function
is that of an ensemble, and an observable is thought to be an
objective property of the microscopic object. Bohr departed from
Einstein's interpretation of an observable only by replacing an
objectivistically realist interpretation by a contextualistically
realist one. Admittedly, Bohr's instrumentalist interpretation of
the wave function, as well as his reference to ``the possible
types of predictions regarding the future behavior of the system''
in his answer \cite{Bohr35} to EPR sound vaguely empiricist.
However, because Bohr, while leaning heavily on the strong form of
the correspondence principle, restricted his considerations mainly
to observables (physical quantities), this empiricism, if existing
at all, did not carry much weight. In particular, his failure to
notice the difference between EPR and EPR-Bell experiments
demonstrates that for Bohr, like for Einstein, a quantum
mechanical measurement result is a property of the microscopic
object rather than a property of a macroscopic measuring
instrument.

In many textbooks of quantum mechanics the way the state vector is
dealt with can also hardly be distinguished from a realist one,
even when adherence to the Copenhagen (orthodox) interpretation is
acknowledged. The paradoxes stemming from this inclination towards
realism have plagued the quantum mechanical literature over the
years, and it has been realized many times that a more empiricist
approach could solve most, if not all, problems (e.g. Wheeler
\cite{Wheeler79}). On the other hand, it has been the Copenhagen
reference to `measurement' in the interpretation of quantum
mechanics, which has been a bone of contention to those believing
that quantum mechanics is applicable outside the context of a
measurement. This is worded in a fairly dramatic way, for
instance, by Popper (\cite{Pop}, p.~2): ``It seems to me that the
attack on realism, though intellectually interesting and
important, is quite unacceptable, especially after two world wars
and the real suffering - avoidable suffering - that was wantonly
produced by them; and that any argument against realism which is
based on modern atomic theory - on quantum mechanics - ought to be
silenced by the memory of the reality for the events of Hiroshima
and Nagasaki.'' It seems to me, however, that judgments like
Popper's one given above are based on an identification of
empiricism and anti-realism, which is hardly defensible in
general. It does not seem reasonable to treat a modest assessment
of the meaning of a physical theory (as merely describing certain
observed aspects of reality) as if this view would deny all
reality behind the phenomena.

Perhaps Popper can be excused because, due to the idea of
`completeness of quantum mechanics in a wider sense' it could be
thought that the Copenhagen interpretation endorses such an
anti-realist view. However, as was seen in section~\ref{sec2.1.1},
such a contention can hardly be attributed to the Copenhagen
interpretation because it is rather `completeness of quantum
mechanics in a restricted sense' that is characteristic of this
interpretation. Whether this latter concept is applicable to the
violent events that took place in Hiroshima and Nagasaki, depends
on the question of what is the domain of application of quantum
mechanics. Application of quantum mechanics to macroscopic events
may be possible to a certain extent, but we should be aware of the
possibility that such events may transcend the domain of
application of the theory (even though the theory may work well
for a description of certain aspects of the microscopic processes
that are involved). For a description of microscopic events
quantum mechanics may play an analogous role to the one played by
the theory of rigid bodies in describing solids: the theory may
describe certain features of reality as they are under certain
conditions defining the domain of application of the theory; under
extreme conditions the theory may become inapplicable, and new
theories will be needed to describe features of reality not
covered by it. Unfortunately, in Popper's philosophy of science a
theory's domain of application is largely absent.

Virtually all paradoxes that over the years have plagued quantum
mechanics stem from a realist interpretation. Sometimes it is
possible to alleviate a problem by relaxing from an
individual-particle interpretation to an ensemble one, thus
assuming, like in Schr\"odinger's cat paradox, that the wave
function does not describe the reality of a single cat but of an
ensemble of cats. However, the persistence of the so-called `cross
terms' in the density operator keeps provoking a necessity of
appealing to `observation' in order to account for their
unobservability (e.g. Jauch \cite{Jauch}, chapter~11). In contrast
to Einstein's contention, an ensemble interpretation of the wave
function does not solve all problems. The EPR problem, discussed
in section~\ref{sec2.4}, is a case in point here. An
objectivistic-realist interpretation of conditional preparation of
the particle $2$ ensemble by a measurement on particle $1$ keeps
being problematic (although less acutely than in an
individual-particle interpretation) because it attributes, for
each member of the ensemble of particle pairs, to particle $2$
sharp values of both incompatible observables $B^{(s)}_2$ and
$B'_2$. An empiricist interpretation offers an alternative to the
nonlocality of Bohr's contextualistic realism by interpreting a
transition from state vector $|\psi_{12}\rangle$ (\ref{3}) to the
state vector $|\beta^{(s)}_{2i}\rangle$ not as a transition to a
description of a subensemble of the ensemble of particles $2$, but
as a transition to a new preparation procedure for particles $2$,
in which the latter particles are selected on the basis of the
measurement results read off the measuring instrument for particle
$1$. Since such a preparation procedure is manifestly causal the
nonlocality problem does not arise.

It is important to note that an empiricist interpretation of EPR
in this vein is not liable to the problem facing Einstein's
solution on the basis of an objectivistic-realist ensemble
interpretation, to the effect that according to the latter
interpretation well-defined values of incompatible observables
would be attributable to each particle of the ensemble, thus
causing the problems of the Kochen-Specker and Bell type referred
to in sections~\ref{sec2.1} and \ref{sec2.5}. In an empiricist
interpretation in the EPR experiment particle $2$ is not thought
to acquire a well-defined value of $B^{(s)}_2$ as soon as a value
of $A^{(s)}_1$ is ascertained. Such a value is obtained only if
the observable is actually measured (as is the case in an EPR-Bell
experiment). Hence, in a way an empiricist interpretation
subscribes to Jordan's view that a quantum mechanical measurement
result is created during a measurement. However, it is not created
as a property of the microscopic object, but as a property of the
measuring instrument (pointer position), which, for this reason,
should be actually present. It should be noted that procedures
like conditional preparation of particle $2$ in the EPR experiment
are not at all exceptional. They are quite common experimental
procedures for preparing microscopic objects ``in certain
well-defined states'' (for instance, the Compton-Simon
experiment). Note, however, that in an empiricist interpretation
the realist terminology, in which a state vector is thought to
represent the {\em result of a preparation}, should be replaced to
the effect that the wave function is rather representing the
preparation itself. For describing states of the microscopic
objects we have to take recourse to subquantum theories.

The neo-Copenhagen interpretation developed in the present paper
endorses an empiricist interpretation of the quantum mechanical
formalism. By thus weakening the interpretation paradoxes can be
evaded. Far from ``restrict(ing) quantum mechanics to be
exclusively about piddling laboratory operations'' this allows
application of the theory to any preparation procedure
representable by a quantum mechanical state vector or density
operator: it is not necessary that the preparation be man-made.
However, things are different with respect to observables
(represented by Hermitian operators or POVMs). Quantum mechanics
is intended in the first place to describe observations made by
means of measuring instruments devised especially so as to create macroscopic
phenomena, the relative frequencies of which can be registered
conditionally on a certain preparation procedure. In the
neo-Copenhagen interpretation quantum mechanics is thought just to
describe relations between preparations and such measurement
phenomena, which, however, are {\em not} to be interpreted in
Bohr's realist sense but in the sense of an empiricist
interpretation. Quantum mechanical measurement results refer to
pointer positions of quantum mechanical measuring instruments.
What is a quantum mechanical measuring instrument cannot be
defined independently of the theory (i.e. quantum mechanics)
itself. It is a matter of experience whether a certain
experimental procedure is a valid measurement procedure within the
domain of application of quantum mechanics. In particular, this
domain has been appreciably extended as a consequence of the
discovery that Hermitian operators do not exhaust all
possibilities of representing quantum mechanical measurements.
\section{Subquantum theories}\label{sec4}
An interesting question, relevant to the neo-Copenhagen
interpretation because it denies to quantum mechanics
`completeness in a wider sense', is a characterization of that
theory's domain of application. As yet, we do not have any
experimental clue helping us to answer this question, and
necessary for knowing what kind of theories will be suitable for
transcending quantum mechanics. Most attempts at devising
subquantum theories had the intention to restore a classical,
often deterministic, view (e.g. Bohm \cite{Bohm52}). We do not
have any reason, however, to think that at a submicroscopic level
the world will be more similar to our macroscopic one than it is
at the microscopic level.

Whereas the Copenhagen interpretation had the opportunity to
neglect questions with respect to subquantum theories on the basis
of a completeness claim of quantum mechanics (even though
ill-understood), the neo-Copenhagen interpretation has to take
these questions seriously. In particular, it has to cope with an
alleged nonlocality, induced by derivations of the Bell
inequalities from subquantum theories, even after it has been
possible to banish the ghost of nonlocality at the quantum level
by switching from a realist to an empiricist interpretation.
However, as will be argued in the following, the nonlocality
claim, based on violation of the Bell inequalities by certain
EPR-Bell experiments, actually is a consequence of a completeness
claim of the subquantum theory used, in the sense that it is
assumed that this theory will describe all possible measurements,
including quantum mechanical ones. It is not necessary to conclude
that nonlocality is the cause of this violation if in the theory
additional assumptions are made that are not satisfied in reality,
and, therefore, make the theory inapplicable even if it is local.
It is not even plausible that in case of quantum mechanical
measurements nonlocality is the cause, because the possibility of
violating the Bell inequalities by quantum mechanical measurements
hinges on {\em in}compatibility of quantum mechanical observables,
which, due to the (unchallenged) postulate of local commutativity,
is a {\em local} affair. It is very well possible that the
classical paradigm, referred to above, is such an additional
assumption. In the following it is demonstrated how by getting rid
of this paradigm also at the level of subquantum theories the
ghost of nonlocality can be driven off.

Let us first briefly review derivations of the Bell inequalities
from hidden variables theory. Allegedly (\cite{Eber78}) the most
general one has been given by Clauser and Horne \cite{ClH74}. It
is based on the following representation of the quantum mechanical
probability $p_i$ of measurement result $a_i$ of quantum
mechanical observable $A$:
\begin{equation}\label{4.1}
p_i=\int_\Lambda d\lambda\;\rho(\lambda)\;p_A(a_i|\lambda).
\end{equation}
Here $\lambda$ is a submicroscopic (hidden) variable, to be
compared with the phase space point $(q,p)$ of classical
mechanics, phase space being generalized to the hidden variables
space $\Lambda$. The probability of $\lambda$ is given by
$\rho(\lambda)$. The quantity $p_A(a_i|\lambda)$ is the
conditional probability of measurement result $a_i$ for given
$\lambda$. It actually is a representation of the detection
process. If $p_A(a_i|\lambda)$ can have only values $0$ and $1$
the theory is called deterministic (this might implement the idea
of `faithful measurement', which does not have a meaning in an
empiricist interpretation of quantum mechanics). It should be
realized that this determinism refers only to the detection
process; the free evolution in phase space, governing the time
dependence of $\rho(\lambda)$, may be either deterministic or
indeterministic. Since $p_A(a_i|\lambda)$ may depend on the
measurement procedure used for measuring observable $A$, this
theory can even account for contextuality of quantum mechanical
observables. For EPR-Bell experiments (\ref{4.1}) is generalized
to
\begin{equation}\label{4.2}
  p_{ij}=\int_\Lambda d\lambda\;\rho(\lambda)\;p_{A_1A_2}(a_{1i},a_{2j}|\lambda),\;
  p_{A_1A_2}(a_{1i},a_{2j}|\lambda)=p_{A_1}(a_{1i}|\lambda)p_{A_2}(a_{2j}|\lambda),
\end{equation}
in which the assumption of locality is implemented by a condition
of conditional statistical independence of the bivariate
conditional probabilities $p_{A_1A_2}(a_{1i},a_{2j}|\lambda)$, the
conditional probability of observable $A$ being independent of $B$
(and vice versa). A simple way to prove that the detection
probabilities of the pairs of observables
$(A_1,A_2),\;(A_1,B_2),\;(B_1,A_2),$ and $(B_1,B_2)$ satisfy the
Bell inequalities (de Muynck et al. \cite{dMDBMa95}) relies on the
possibility to construct, on the basis of the validity of
representation (\ref{4.2}) for each of these measurements, the
quadrivariate probability distribution
\begin{equation}\label{4.2.2}
p_{ijk\ell}=\int_\Lambda d\lambda\;
\rho(\lambda)\;p_{A_1}(a_{1i}|\lambda)\;p_{B_1}(b_{1j}|\lambda)\;
p_{A_2}(a_{2k}|\lambda)\;p_{B_2}(b_{2\ell}|\lambda).
\end{equation}

The widely accepted idea that the quantum world must be nonlocal
is based on the assumption that locality is the {\em only}
presupposition in deriving the Bell inequalities from this theory,
thus enabling to pinpoint nonlocality as the only possible cause
of their violation in EPR-Bell measurements. Yet, there is still
another assumption involved in the representations (\ref{4.1}) and
(\ref{4.2}) of quantum mechanical detection probabilities. It is
important to note that in these expressions by a conditional
probability like $p_A(a_i|\lambda)$ quantum mechanical measurement
results $a_i$ are conditioned on an {\em instantaneous} value of
hidden variable $\lambda$. Is it reasonable to assume that a
quantum mechanical measurement result $a_i$ is determined, even in
a stochastic sense, by such an instantaneous value? If so, what is
the precise instant of time at which the value of $\lambda$ should
be taken in the conditional probabilities? Given the possibility
that $\lambda$ is fluctuating very fast even compared to the
characteristic transition times of quantum phenomena, it is
probable that a quantum mechanical measurement does not probe an
instantaneous value of $\lambda$ at all, but (a part of) a {\em
trajectory} $\bar{\lambda}$, much in the same way a
(thermodynamic) measurement of temperature or pressure does not
probe an instantaneous value of $(q,p)$ in classical statistical
thermodynamics. Subquantum theories of the types in which quantum
mechanical measurement results are conditioned on $\lambda\;
(\bar{\lambda})$ are referred to as quasi-objectivistic
(non-quasi-objectivistic) subquantum theories in de Muynck
\cite{dM2002}.

The analogy between quantum mechanics and statistical
thermodynamics has been drawn many times before (e.g. de Broglie
\cite{deBroglie95}, Bohm et al. \cite{Bohm53,BoVi}, Nelson
\cite{Nelson67,Nelson85}, Davidson \cite{Dav79}, D\"urr et al.
\cite{DuGoZa92}). If the analogy is sound, this implies that
quantum mechanical measurements may not be fast enough to probe
subquantum fluctuations. This means that for application to
quantum mechanical measurements the conditional probabilities
$p_A(a_i|\lambda)$ should probably be replaced by
$p_A(a_i|\bar{\lambda})$, and the expression (\ref{4.1}) by a
density functional like
\begin{equation}
p_i=\int_{\bar{\Lambda}}
d\bar{\lambda}\;\rho(\bar{\lambda})\;p_A(a_i|\bar{\lambda}),\label{4.3}
\end{equation}
in which $\bar{\Lambda}$ is a space of possible trajectories
$\bar{\lambda}$, and $\rho(\bar{\lambda})$ is a probability
distribution of these trajectories. The difference between
$\lambda$ and $\bar{\lambda}$ marks the distinction between
subquantum elements of physical reality and quantum mechanical
ones, referred to in section~\ref{sec2.1.3}.

By itself a transition from (\ref{4.1}) to (\ref{4.3}) is not
sufficient to block derivation of the Bell inequalities, since in
(\ref{4.2.2}), too, $\lambda$ could be replaced by
$\bar{\lambda}$, thus allowing the construction of a quadrivariate
probability distribution even if all conditional probabilities are
conditioned on trajectories. In order to prevent such a
construction it is useful to remember that, as far as quantum
mechanical measurements probe reality, they probe a {\em
contextual} reality (cf. section~\ref{sec2.7.1}). In a subquantum
theory describing such measurements this contextuality might be
implemented by a contextuality of the trajectory $\bar{\lambda}$,
which could be co-determined by the experimental arrangement in
the same way as the canonical state $Z^{-1}e^{-H/kT}$ of classical
statistical thermodynamics is co-determined by it (for instance,
by the shape of the container of a gas). For this reason, in the
context of a measurement of quantum mechanical observable $A$ in
(\ref{4.3}) the trajectory $\bar{\lambda}$ has to be replaced by a
trajectory $\bar{\lambda}^A$ that is dependent on the measurement
arrangement. Hence, instead of (\ref{4.3}) we get
\begin{equation}\label{4.3.2}
p_i=\int_{\bar{\Lambda}^A}
d\bar{\lambda}^A\;\rho(\bar{\lambda}^A)\;p_A(a_i|\bar{\lambda}^A),
\end{equation}
in which $\bar{\Lambda}^A$ is the space of trajectories allowed
within the context of a measurement of observable $A$.

Contextuality implies that it is natural to assume that for
incompatible observables $A$ and $B$ (characterized by mutually
exclusive measurement arrangements) we in general have
\begin{equation}\label{4.4}
\bar{\lambda}^A \neq \bar{\lambda}^B.
\end{equation}
Hence, even if in different experiments the object were prepared
in the same hidden variable state $\lambda$, the trajectories are
distinct if the measurement arrangements are mutually exclusive.
For incompatible observables we also have
\[\bar{\Lambda}^A \neq \bar{\Lambda}^B.\]
Now the important point is that in the representations
(\ref{4.3.2}) of probability distributions of incompatible
observables the conditional probabilities cannot be conditioned on
the same states (trajectories). As a consequence, the
construction, analogous to (\ref{4.2.2}), of a quadrivariate
probability distribution fails, and the Bell inequalities cannot
be derived any more (even though these may be satisfied for
measurements in the subquantum domain, fast enough to probe the
instantaneous value of $\lambda$).

The assumption that quantum mechanical measurements do not probe
the instantaneous value of $\lambda$, but only (contextual)
trajectories $\bar{\lambda}^A$ is sufficient to break the
unnatural connection widely supposed to exist between nonlocality
and violation of the Bell inequalities. Derivation of the Bell
inequalities for EPR-Bell experiments like those performed by
Aspect et al. \cite{Asp81,Asp82} is impossible, then, because an
object cannot be in the same state (trajectory) for all of the
four different measurements involved. This explanation is in
agreement with the fact that {\em in}compatibility is a necessary
condition for the Bell inequalities to be violated. Since in an
EPR-Bell experiment the trajectory of each of the particles of the
particle pair can be thought to be {\em locally} co-determined by
its own measurement arrangement, the solution proposed here for
subquantum theory is essentially the same as the one proposed for
quantum mechanics (cf. section~\ref{sec2.5.1}). No nonlocal
influences are necessary to violate the Bell inequalities.

It is interesting to consider the possibility that an individual
preparation within the context of a measurement of observable $A$
might be describable by a trajectory $\bar{\lambda}^{(A,a_i)}$,
yielding with certainty measurement result $a_i$ if the
measurement is a faithful one. This would restore determinism of
the measurement process, to the effect that a measurement result
$a_i$ refers in a deterministic way to a subquantum state, be it a
trajectory. A connection between quantum mechanics and the
subquantum theory proposed here might be provided by the
contextual state $\rho_A$ (\ref{2.7.1.1}), which may be referring
to the preparation of the subquantum states (trajectories)
$\bar{\lambda}^{(A,a_i)}$ in an ensemble symbolically represented
by the (statistical) state $\bar{\lambda}^A$, relative frequencies
being given by (\ref{2.7.1.1}). In general
$\bar{\lambda}^{(A,a_i)}\neq \bar{\lambda}^{(B,b_j)}$ if $A$ and
$B$ are incompatible.

Finally, it is interesting to note that the solution presented
here corroborates a certain nonlocality of quantum mechanics (e.g.
de Muynck \cite{dM84}) quite different from the nonlocality
involved in the usual explanation of violation of the Bell
inequalities. This nonlocality has to do with the domain of
application of a theory, delimiting the kind of objects described
by it. Thus, an electron is not a point particle, but it is an
extended object. However, within the domain of application of
quantum mechanics it is to be considered as a
(nonlocal/inseparable) whole, in the same way a billiard ball in
the theory of rigid bodies, or a volume of gas in a thermodynamic
equilibrium state is to be considered an inseparable whole. This
does not imply that the objects are really nonlocal objects, in
the interior of which nonlocal interactions would be responsible
for inseparability, or for relative immobility of its parts. What
it means is that within the domain of application of the relevant
theory the objects can be considered as primitive entities. When
experimentally leaving the domain of application of the theory the
apparent nonlocality may disappear, analogously to the
impossibility of staying within the domain of application of rigid
body theory under experimental conditions capable of splitting a
billiard ball.

\section{Summary}
\label{sec5} In this paper an interpretation of the mathematical
formalism of quantum mechanics is proposed, remedying the
confusions and inconsistencies of the Copenhagen interpretation
while maintaining the essential role attributed by the latter
interpretation to the interaction of microscopic object and
measuring instrument. For this latter reason it is proposed to
refer to the new interpretation as a neo-Copenhagen one. It has
particularly been inspired by the recent insight that, in order to
encompass all possible measurements within the domain of quantum
mechanics, it is necessary to generalize the mathematical
representation of quantum mechanical observables to positive
operator-valued measures, thus making obsolete the preferred
position Hermitian operators have in the Copenhagen interpretation
(as well as in most other interpretations). Thus, von Neumann's
projection postulate is not applicable to generalized observables.
Another reason to refer to the Copenhagen interpretation is that the
notion of complementarity as a consequence of mutual disturbance
in a joint measurement of incompatible observables remains one of
the cornerstones of the new interpretation.

In the neo-Copenhagen interpretation the empiricist tendencies, to
be observed in the Copenhagen interpretation but not implemented
there in a consistent way, are taken seriously. By adopting an
empiricist interpretation in which the mathematical formalism is
taken to refer to macroscopic procedures of preparation and
measurement rather than to the microscopic object, the paradoxes
haunting quantum mechanics can be solved. In particular, the
nonlocality problem, induced by the EPR experiment, does not exist
any more. The contextual meaning of quantum mechanics, underlined
by Bohr's analysis of the EPR proposal, is corroborated, but at
the same time corrected by drawing a distinction between
measurement and preparation, largely neglected in the Copenhagen
interpretation. Bohr's instrumentalist interpretation of the wave
function as well as Einstein's realist ensemble interpretation are
shown to be wanting.

The discussion on the issue of the completeness of quantum
mechanics is shown to be very confusing because two different
notions of completeness are at stake. It is argued that an
empiricist interpretation of quantum mechanics leaves room for
subquantum theories. The concomitant nonlocality problem induced
by the violation of the Bell inequalities is analyzed, and shown
to be a consequence of a too restricted subquantum theory,
unjustifiedly conditioning quantum mechanical measurement results
on instantaneous values of the subquantum (hidden) variables. A
comparison of the relation between quantum mechanics and such
(quasi-objectivistic) subquantum theories with similar relations
between theories describing physical phenomena at different levels
of observation, demonstrates that presumably the assumption of
quasi-objectivity cannot be maintained. Contrary to Bell's
contention that nonlocality is the essential assumption allowing
to derive the Bell inequalities, it is rather the assumption of
quasi-objectivity (overlooked by Bell) that may be responsible.\\


\noindent {\bf \large Acknowledgment}

\noindent The author thanks professor N.G. van Kampen and Peter
Morgan for valuable remarks.


\end{document}